# Turning Errors into Successes: Designing LLM Tutoring Around Scaffolded Error Correction

Michael W. Asher[1], Meng Cao[1], and Paulo F. Carvalho[1]

[1]Human-Computer Interaction Institute, Carnegie Mellon University, Pittsburgh, PA

November 15, 2026

## Author Note

Correspondence concerning this article should be addressed to Michael W. Asher, Human Computer Interaction Institute, Carnegie Mellon University, Newell-Simon Hall, 5000 Forbes Avenue, Pittsburgh, PA 15213, United States. Email: masher@andrew.cmu.edu.

This research was supported in part by National Science Foundation grants #2541106 and #2301130 to PFC, a Carnegie Mellon University Seed Grant for Generative AI and Education Research to PFC and MWA, and a Learning Engineering Institute Seed Grant from Carnegie Mellon University to PFC, MWA, and MC. All data and code for this study are openly available at https://github.com/the-oak-lab/scaffolded-error-correction. Methods, hypotheses, and analyses were preregistered at https://osf.io/vfa6w.

**Abstract**

Large language models (LLMs) could make personalized tutoring available at scale, but it remains unclear how they can best support the instructional processes that make one-on-one instruction effective. We compared two ways of using LLMs after students make errors: providing personalized explanations of those errors, or engaging students in additional cycles of scaffolded practice and feedback, with a Socratic tutor guiding students toward the correct answer. In a preregistered online experiment, 400 adults practiced setting up constraint equations and, after each attempt, received the correct answer, a generic explanation, a personalized LLM explanation of their error, or Socratic tutoring. Socratic tutoring helped: participants resolved 64% of their errors through the dialogue and scored 0.24 SD higher on practice-aligned posttest questions than those who received generic or personalized explanations. Socratic participants also found the session more interesting, less frustrating, and less costly, even though it took nearly twice as long. In contrast, we found no evidence of a benefit of personalized explanations over generic ones, and neither type of explanation reliably outperformed simply seeing the correct answer. These findings suggest that LLM tutors may add the most value not by explaining errors more precisely but by helping students correct those errors themselves through scaffolded practice. This pattern fits a mastery-experience account of tutoring, in which an effective tutor notices when a student is struggling, steps in with guidance, and works with the student until the struggle resolves into success. AI tutors may be most effective when designed around this principle.

# 1. Introduction

One-on-one tutoring is among the most powerful interventions for struggling students (Dietrichson et al., 2017; Nickow et al., 2024). A skilled tutor can diagnose the gaps in a student's understanding and provide precisely tailored practice and feedback to close them. The problem is that human tutoring is impractical and expensive at scale, putting it out of reach for most. For forty years, educational technologists have tried to build tutoring programs that offer this kind of personalized instruction, but they have been limited by an authoring problem: developers must anticipate every possible mistake to pre-write personalized hints and feedback (Aleven et al., 2016; Dermeval et al., 2018).

Recent breakthroughs in large language models (LLMs) appear to offer a way around this personalization bottleneck, and developers have moved quickly to build generative AI tutors that promise fully adaptive one-on-one instruction at scale. So far, however, the evidence for LLM tutors has been discouraging. The most straightforward approach is to give students a general-purpose chatbot prepared with the task they are attempting and a short prompt to help. This tends to backfire because students often seek help by simply asking for answers, bypassing the productive struggle required for learning, and learning less as a result (Bastani et al., 2025; Kreijkes et al., 2026; Stadler et al., 2024).

A common fix has been to add safeguards that instruct the LLM to withhold answers and respond with hints or scaffolded, Socratic, questions. Rigorous evaluations of such tutors are scarce, but the existing evidence about their efficacy is modest at best. In a two-year randomized trial of Khanmigo in 18 middle schools, Oreopoulos & Low (2026) embedded the tutor in daily remedial math sessions. Nearly every student tried it, but the median student messaged Khanmigo on only a third of the days they practiced, engaged with it in only 17% of the sessions in which they made errors, and only 14.5% of their messages contained a mathematical question or step of reasoning. Effects on standardized test scores were small (0.06–0.08 SD per year), roughly matching Khan Academy practice without generative AI (approximately 0.10 SD at the same level of use; Eames et al., 2026) and falling far short of human tutoring (0.32 SD in a meta-analysis of 36 studies; Dietrichson et al., 2017). A randomized experiment with high school math students found a similar result for a GPT-4 tutor with hint-based guardrails: the guardrails eliminated the drop in learning caused by an unrestricted chatbot, but students learned no more than peers who practiced with a textbook and notes (Bastani et al., 2025).

The gap between the effectiveness of human and generative AI tutoring raises an important question about what makes tutoring effective. Human tutoring combines multiple instructional processes that support learning and motivation: a skilled tutor diagnoses gaps in students' understanding, provides personalized feedback, elicits active responses, and adapts their instruction based on those responses. How can LLM tutoring be designed to support these processes? We consider two approaches: (1) personalized explanations of students' errors, and (2) interactive Socratic tutoring that intervenes after each mistake to scaffold students toward a correct response.

### 1.1. Personalized Explanations of Students' Errors

Of the processes that occur in one-on-one tutoring, it may be the most feasible for LLMs to focus on personalized feedback. The others require sustained interaction: eliciting responses and adapting to them makes tutoring a multi-turn problem in which each move depends on an accurate read of what the student understands (VanLehn et al., 2007), something the tutor cannot observe directly. The tutor must infer misconceptions from ambiguous input, and students often give it little to work with. In the Khanmigo trial, most student messages were one- or two-word answers or clicks on suggested prompts (Oreopoulos & Low, 2026). The tutor must then give hints specific enough to move the student forward but not so specific that they give away the answer, and it must change strategy when a student stalls or grows frustrated. Models that misjudge knowledge gaps or miscalibrate their questions may overwhelm struggling learners, replacing one unanswerable question with a series of them (Kirschner et al., 2006) and producing frustration and disengagement (Ryan & Deci, 2000).

Explaining an error, by contrast, is a single-turn problem: given the task, the correct answer, and the student's response, the model only needs to write a clear explanation. LLMs can already generate explanatory feedback that improves learning relative to no feedback (Liebenow et al., 2025; Meyer et al., 2024), and they can do so on the fly, without developers pre-writing responses to every anticipated mistake. The open question is whether personalized explanations add anything beyond the generic ones typically embedded in tutoring systems.

There is a case to be made for personalized feedback. Explicit guidance is an important part of instruction (Kirschner et al., 2006), and an explanation written for a student's particular error can address the misconception behind it directly. A generic explanation leaves it to students to work out which parts apply to them, a diagnostic task novices may be poorly placed to perform (Sweller, 2004). A personalized explanation may spare these learners the cognitive load of mapping a general account onto a specific error, making explicit the contrast between what they did and what was needed.

There is also a case against it: students learn most effectively from a cycle of practice, errors, and feedback (Asher et al., 2025; Asher & Carvalho, 2026; Bloom, 1968; Carpenter et al., 2022; Ericsson et al., 1993; Freeman et al., 2014; Koedinger et al., 2015; Macnamara et al., 2014; Pan & Carpenter, 2023; Roediger & Karpicke, 2006). Information delivered passively through an upfront explanation tends to wash over students without engaging their cognitive resources (Carvalho et al., 2022, 2024). Practice changes this by letting students make errors that surface gaps in their understanding (Bjork et al., 2013; Pan & Carpenter, 2023), showing them what they have mastered and where they were overconfident. Feedback then has a clear job: resolving those errors. Because students know what they are missing, they can focus on it and encode it more thoroughly (Asher et al., 2025; Asher & Carvalho, 2026; Gold et al., 2026). If practice prepares students to find what they need in a generic explanation, personalization may add little.

### 1.2. Socratic Feedback

The case against personalization points to a different role for a tutor. If students learn best from a cycle of practice, errors, and feedback, a tutor's value may lie less in explaining errors than in eliciting responses and adapting to them. This suggests a second approach: Socratic feedback, in which an LLM tutor guides students to correct their errors themselves. This is a version of the harder, multi-turn problem described above, but if modern LLMs prove capable of diagnosing errors and scaffolding questions, it could be a powerful approach for scalable tutoring.

Generative AI tutors, as they are typically designed, do little to ensure that students resolve their errors. These tutors are usually reactive: chatbots that sit in a box on the side of the page, waiting to be asked for help. Because these tutors are always available, students can seek help before attempting a problem, bypassing some of the practice that makes instruction effective. And because they respond only when asked, students who make errors will receive personalized feedback only if they think to request it. This is especially costly because students in educational technology often abandon assignments after mistakes (Arztmann et al., 2025;

Botelho et al., 2019; Johansson et al., 2024), so a tutor that waits may never get the chance to help. In the Khanmigo trial, for example, students engaged with the tutor in only 17% of the sessions in which they made errors (Oreopoulos & Low, 2026).

Whether errors get resolved also matters motivationally. Success and failure send strong signals about competence and potential, beliefs that are central to major theories of motivation (Ryan & Deci, 2000; Wigfield & Eccles, 2000). When errors go unresolved, students may take them as evidence that they lack the ability to succeed. However, errors can also open the door to mastery experiences, in which a student overcomes a challenge and learns that they are capable of doing so, building the hard-earned self-efficacy that sustains future effort (Bandura, 1997).

If these barriers are limiting what students gain from LLM tutors, a Socratic tutor built into the feedback cycle might deliver where chatbots on the side of the page have not. By requiring an attempt first, stepping in after each error, and returning students to practice until they solve the problem, a proactive tutor could help students complete a cycle of practice, errors, feedback, and ultimate success, turning mistakes into mastery experiences.

## 2. The Present Research

In the present research, we compared these two approaches for LLM tutoring: personalized explanations of students' errors and interactive Socratic tutoring built into the feedback cycle. In a preregistered experiment, 400 adults practiced setting up constraint equations in an online tutoring session. After each attempt, they received one of four types of feedback.

Two conditions provided benchmarks. In the correct-answer condition, participants saw only whether they were correct and the correct equation. Because students can learn from cycles of practice, errors, and even simple corrections (Roediger & Karpicke, 2006), this condition is a meaningful baseline: it should produce substantial learning on its own, and it shows what each additional form of support contributes. In the generic explanation condition, participants saw a general account of why the correct equation is right. This is a standard approach in educational technology (Van der Kleij et al., 2015) and the benchmark that LLM-based support should beat.

The other two conditions tested the LLM approaches. In the personalized explanation condition, an LLM supplemented the generic explanation with an account of the participant's specific error, testing whether the simpler task of personalized feedback adds value beyond a good generic explanation. In the Socratic tutoring condition, an LLM tutor stepped in after each error and guided participants toward the correct answer with scaffolded questions. This condition tested whether building interactive tutoring into the cycle of practice, errors, and feedback improves learning and motivation beyond what explanations provide.

**2.1. Hypotheses**

Personalized explanations and Socratic tutoring give students different opportunities to learn from their mistakes. A personalized explanation is an opportunity to learn from information: after an error, the student reads an account of what went wrong and can apply it to the next problem. Socratic tutoring adds opportunities to respond within each problem. Each scaffolded question asks students to apply what they know to a smaller piece of the problem and receive feedback on that attempt. A single error therefore becomes a series of practice-and-feedback cycles that, ideally, ends with the student correcting the error themselves.

These extra opportunities could increase learning, but only if they are productive. As discussed in Section 1.1, a tutor whose questions are poorly calibrated may replace one unanswerable question with several. And even if the questions are well-calibrated, every exchange adds time and effort. Because simple practice with feedback is already effective (e.g.,

Asher et al., 2025; Carpenter et al., 2022; Roediger & Karpicke, 2006), our preregistered predictions were skeptical that Socratic tutoring would add enough to justify these costs.

We hypothesized that Socratic tutoring would be inefficient, producing no benefit for learning relative to the two explanation-only conditions (H1a) while taking much more time (H1b). We also predicted that Socratic tutoring would increase the perceived cost of learning (H2) because of the extra time and effort it would require. Finally, we preregistered two predictions about the value of explanations: the conditions that explained the correct answer would outperform correct-answer feedback in learning (H3) and in perceived effectiveness (H4). Given the cases both for and against personalized explanations, we made no directional hypothesis about personalized versus generic explanations.

## 3. Method

The study was administered online via Qualtrics using a between-subjects design. Participants completed five sections in sequence: a constraint-equation pretest, a baseline questionnaire, a tutoring session, an outcome questionnaire, and a posttest. The study was approved by the institutional review board at Carnegie Mellon University and preregistered at https://osf.io/vfa6w.

### 3.1 Participants

We recruited participants online via *Prolific*. Eligible participants were at least 18 years old, lived in the United States, provided informed consent, and could not already set up a basic constraint equation, as determined by a screening question at the start of the study (see Section 3.2). Participants who answered the screening question correctly (470 in total) were paid $1 and did not continue. Those who answered incorrectly continued and were paid $4 for the full session (approximately 20 minutes), including the screening question. As preregistered, data collection continued until 400 participants had completed the session and met two quality criteria; we excluded 10 participants who failed a keystroke-based check for AI-assisted cheating (Asher et al., 2026) and 5 participants who finished the session in less than 6 minutes.

Participants in the final sample ($N = 400$) were randomly assigned to correct-answer feedback ($n = 101$), explanatory feedback ($n = 100$), personalized explanation ($n = 100$), or Socratic tutoring ($n = 99$). The sample was 59% women and 41% men, with a mean age of 39.6 years ($SD = 12.9$). Participants self-reported their race/ethnicity as White (59%), Black (20%), Asian (8%), multiracial (8%), or another group (3%), and 3% did not report.

### 3.2. Constraint Equation Pretest

At the beginning of the study, participants were told that they would answer a math problem to determine their eligibility. They then saw the following question:

> A farmer has 200 feet of fencing and wants to build a rectangular enclosure along a straight river. The river side doesn't need fencing. Write an equation using $x$ and $y$ for the dimensions of the enclosure, given the fencing constraint.

Below this problem, and below every problem in this study, participants were given one box for sketching the scenario and another for writing the equation (see Figure 1). Responses were scored automatically as soon as participants submitted them. For the pretest, an equation was counted as correct if it described the same line as either $x + 2y = 200$ or $2x + y = 200$. The program tested equivalence numerically, so any algebraically equivalent form was accepted. The sketch was saved but not scored, and participants received no feedback on their answer. To encourage honest answers, participants were not told that a correct answer would end their participation. Those who answered correctly were thanked, redirected to *Prolific*, and paid.

Those who answered incorrectly continued, and were told that they would learn to set up constraint equations, the task they had just attempted, before taking a test on the topic.

**Figure 1**

*The Constraint Equation Pretest*

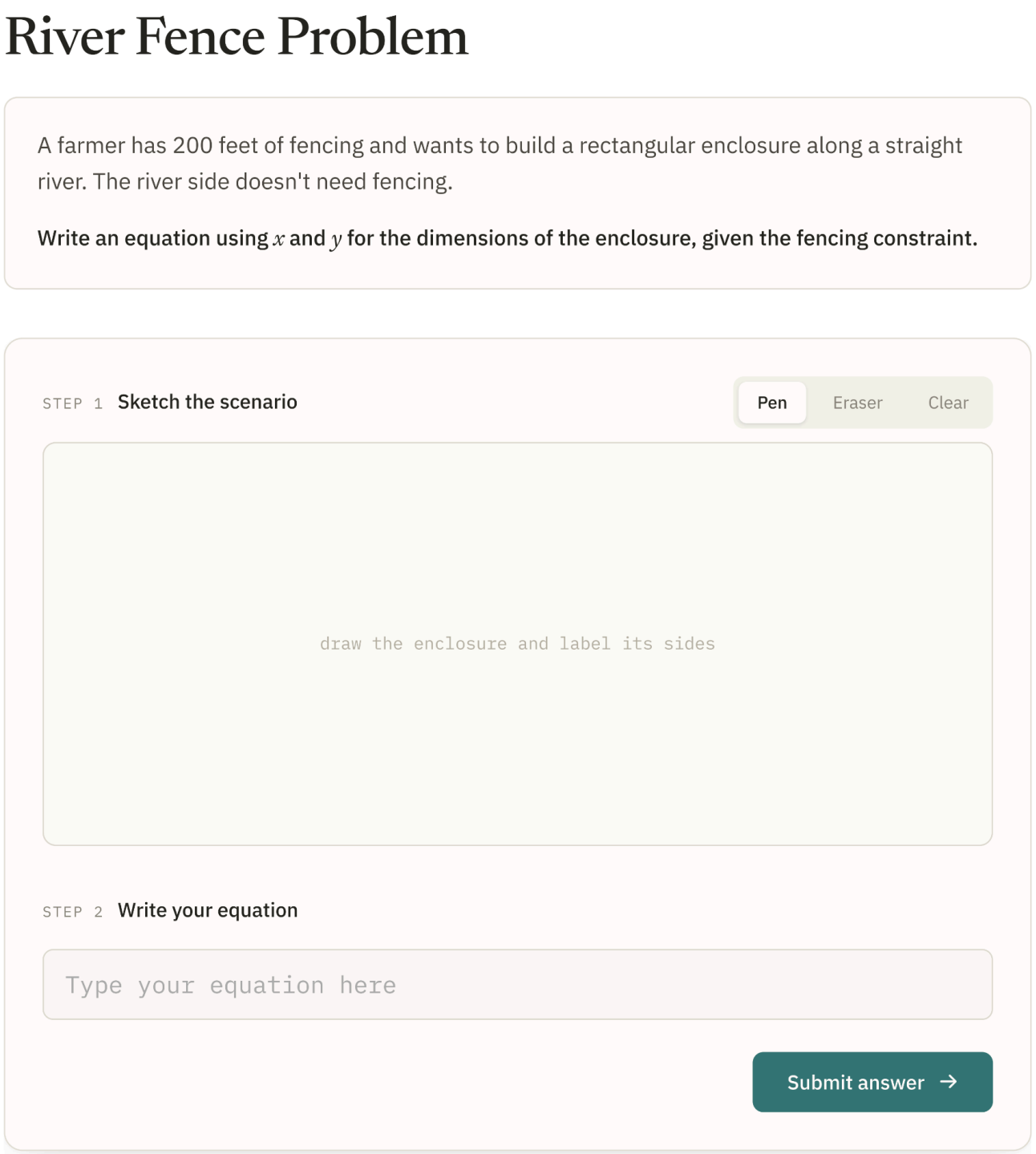

### 3.3 Baseline Questionnaire and Tutoring Session

After the pretest, participants completed a baseline questionnaire with three-item scales measuring their confidence and interest in mathematics. They then began a tutoring session on setting up constraint equations for different types of fences. The session consisted of three practice problems, each with a slightly different learning objective. Whereas the pretest required a constraint equation for a fence with a missing side, the practice session taught participants to model a fence with a gate (Problem 1), a fence with internal dividers (Problem 2), and a fence built from materials with different costs per foot under a fixed budget (Problem 3). Table 1 shows these three problems and their solutions. Every problem had the same two-step format as the pretest: participants first sketched the scenario and then typed an equation. After each submission, participants received feedback according to their randomly assigned condition and then moved on to the next problem. The conditions differed only in the feedback participants received.

**Table 1**

*Constraint Equation Practice Problems*

| Problem Text | Learning Objective | Solution |
|---|---|---|
| A dog daycare is fencing a rectangular play yard using 500 feet of fencing. One of the sides of length $x$ has a 10-foot gate opening that doesn't require fencing, so staff can bring dogs in and out.<br><br>*Write an equation using x and y for the dimensions of the yard, given the fencing constraint.* | Modeling a fence with a gate | $2x + 2y - 10 = 500$ |
| A homeowner is using 1600 feet of fencing to enclose a rectangular backyard. The yard will be divided into 3 equal sections by 2 internal fences running parallel to the sides of length $x$.<br><br>*Write an equation using x and y for the dimensions of the yard, given the fencing constraint.* | Modeling a divided fence | $4x + 2y = 1600$ |
| A rancher is building a rectangular pen. The two sides of length $x$ will use heavy-duty fencing costing \$6 per foot, and the two sides of length $y$ will use standard fencing costing \$2 per foot. The total fencing budget is \$600.<br><br>*Write an equation using x and y for the dimensions of the pen, given the budget constraint.* | Modeling a fence with multiple building costs | $12x + 4y = 600$ |

### 3.4 The Experimental Manipulation: Provision of Feedback

Figure 2 shows an annotated diagram of sample correct-answer feedback, explanatory feedback, and personalized explanation for an incorrect answer.

**Figure 2**

*Elements of Feedback in the Correct-Answer, Explanatory, and Personalized Conditions*

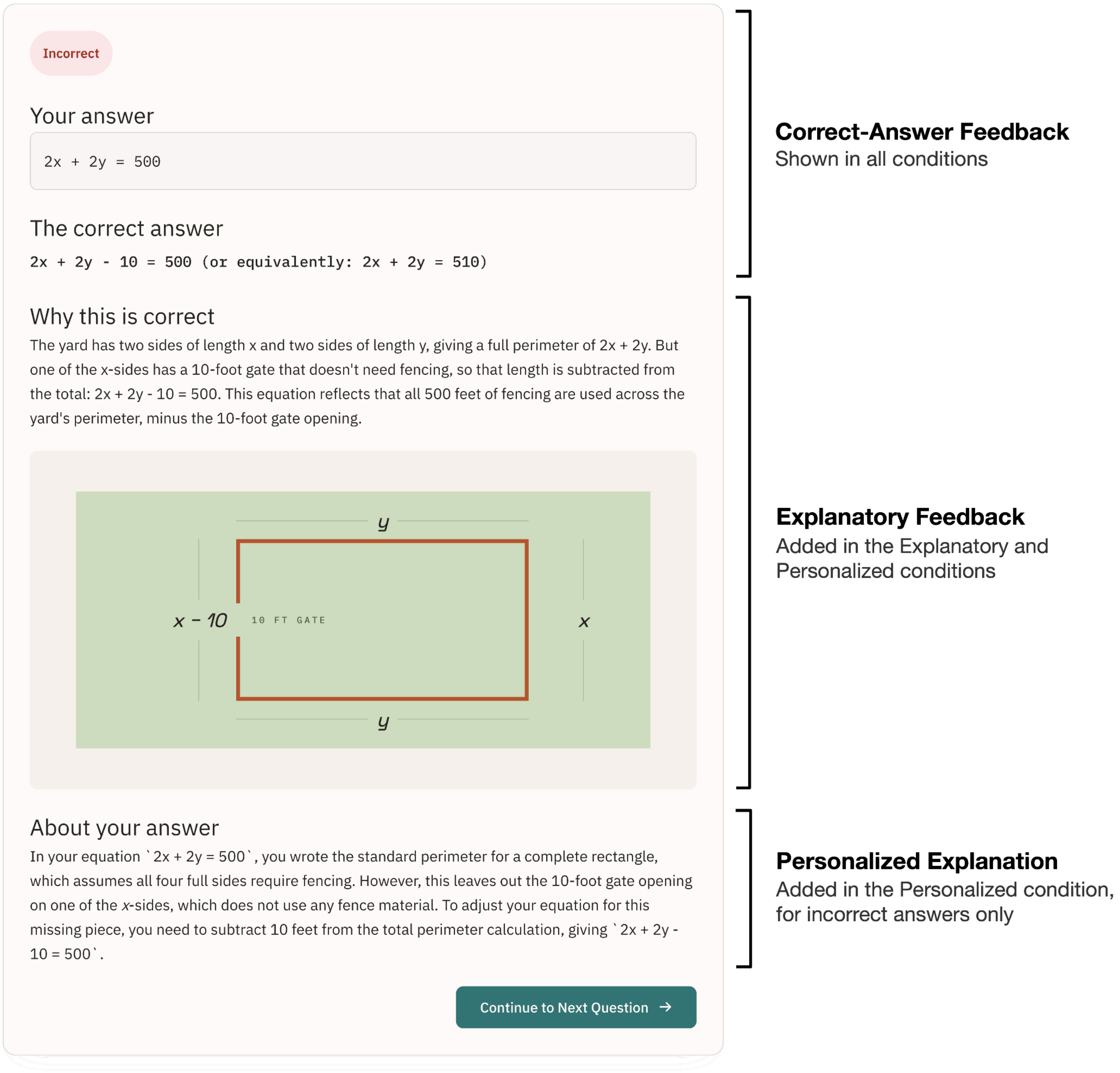

*Note.* The fence diagram was generated by Claude Opus 5, and all explanatory feedback was generated by Gemini 3.5 Flash.

### *3.4.1. Correct-Answer Feedback*

Participants saw whether their equation was correct or incorrect, their own answer, and the correct equation. They received no further explanation.

### *3.4.2. Explanatory Feedback*

Participants saw the same information as in the correct-answer condition, plus a pre-written explanation of why the correct equation is right (generated in advance by Gemini 3.5 Flash) and a labeled diagram of the scenario (generated in advance by Claude Opus 5). This content was identical for every participant regardless of whether their answer was correct, as is typical in educational technology.

### *3.4.3. Personalized Explanation*

Participants who answered correctly saw the same feedback as in the explanatory condition. Participants who answered incorrectly also received a two- to three-sentence, personalized explanation of their specific error (e.g., that they were missing a term or provided an incorrect coefficient). This explanation was generated in real time by Gemini 3.5 flash, based on a prompt that we provide in the Supplementary Material.

### *3.4.4. Socratic Tutoring*

After each submission, participants entered a chat with an LLM tutor (Gemini 3.5 Flash). The tutor was given the problem, the correct equation, and the same pre-written explanation shown in the explanation conditions. It was prompted to withhold the answer and to "use Socratic questioning to guide [the participant] to the correct answer through a series of scaffolded questions" (full prompts are provided in the Supplementary Material). Participants could reply with a revised equation or a question of up to 300 characters per message. When a participant produced a correct equation, either on submission or later in the chat, the tutor confirmed it and briefly explained why it was correct, drawing on the pre-written explanation. If no correct equation had been produced after six tutor replies, the tutor stated the correct equation and explained it. Chats were capped at eight tutor replies or 10,000 total characters, whichever came first. The tutor could also display the labeled diagram of the scenario shown in the explanation conditions if it judged that a participant was struggling to visualize the geometry. If the diagram had not appeared by the end of the chat, it was displayed then, so every participant saw it before moving on. No human reviewed or edited the tutor's messages. Table 2 shows a sample chat with a participant; all chat transcripts are included in the shared data.

**Table 2**

*Sample Socratic Tutoring Session*

| | |
|---|---|
| **Problem** | A dog daycare is fencing a rectangular play yard using 500 feet of fencing. One of the sides of length x has a 10-foot gate opening that doesn't require fencing, so staff can bring dogs in and out. Write an equation using x and y for the dimensions of the yard, given the fencing constraint. |
| **Chat** | **Participant:** x+2y=500<br><br>**Tutor:** You're off to a great start by using the variables *x* and *y*! Let's think about the shape of the yard first. If we were fencing a standard rectangle with sides of length *x* and *y*, and there was no gate at all, what would the expression for the total perimeter be?<br><br>**Participant:** 2x+2y=500<br><br>**Tutor:** Spot on! If there were no gate, we would indeed need $2x + 2y$ feet of fencing to cover the whole perimeter.<br><br>Now, how does having a 10-foot gate, which doesn't require any fencing, affect the total amount of fencing we actually need to put up?<br><br>**Participant:** we don't need that side so I removed one x, leaving the equation x+2y=500<br><br>**Tutor:** I see your thinking! You're definitely on the right track by wanting to subtract the gate's opening from the total fencing.<br><br>Is the gate taking up the *entire* side of length *x*, or is it just a 10-foot section of that side? How would you write the length of that specific side if we only subtract the 10-foot opening?<br><br>**Participant:** Oh, it'd be 2x+2y-10=500<br><br>**Tutor:** That is absolutely correct, fantastic job! Since the full perimeter of the yard is $2x + 2y$, subtracting the ten-foot gate opening perfectly accounts for the fencing we save, giving us the correct equation of $2x + 2y - 10 = 500$. You have successfully set up the equation for the dog daycare!<br><br>[FENCE DRAWING DISPLAYED] |

## 3.5. Final Questionnaire and Posttest

Following the tutoring session, participants completed an outcome questionnaire assessing their experience during the learning session (including their situational interest, distraction, perceived costs, and frustration), the perceived quality of the feedback they received, their interest in learning more about constraint equations, and their confidence and expectations for the posttest.

Immediately after the questionnaire, all participants completed the five-item posttest. Three "practice-aligned" transfer questions were more difficult versions of the three practice problems: one required modeling a fence with two gates, another a fence with both vertical and horizontal dividers, and the third a fence built from three materials with different costs. Two test items went beyond the practiced objectives. A "pretest-aligned" transfer problem involved a fence with a missing side, the same objective tested on the pretest. A "challenge" problem required participants to combine multiple objectives, modeling a fence with two missing sides, different costs per foot, and a gate. Participants received no feedback on the posttest.

## 3.6. Measures

### ***3.6.1. Self-Report and Behavior Measures***

All questionnaire measures used five-point Likert-type scales (ranging from "Not at all ____" to "Extremely ____") or six-point agreement scales ("Strongly disagree" to "Strongly agree"). Measures were adapted from Linnenbrink-Garcia et al. (2010), Durik et al. (2015), Flake et al. (2015), Hecht et al. (2021), and Asher and Harackiewicz (2024). Measures used in the preregistered tests are described here. Additional measures collected for exploratory analyses are described in the Supplementary Material and included in the shared data.

### ***3.6.2. Outcome Measures***

We recorded the time participants spent in the tutoring session (*M* = 8.5 minutes, *SD* = 5.4). As preregistered, we winsorized outliers so that idle participants would not unduly influence the results: for each timer, we set a cutoff at the third quartile plus three times the interquartile range and replaced any longer value with the cutoff. In total, 1.5% of timers were winsorized.

Situational interest in the session was assessed with five items (e.g., "This session has been interesting"; *α* = .94, *M* = 4.3, *SD* = 1.2). Interest in constraint equations was assessed with three items (e.g., "How much would you like to learn more about constraint equations?"; *α* = .95, *M* = 2.9, *SD* = 1.2). Two items measured the perceived cost of the practice session: emotional cost ("The practice session was emotionally draining") and effort cost ("The practice session required too much effort"). As preregistered, we averaged these items into a single index (*α* = .81, *M* = 2.7, *SD* = 1.3). A single item, "The practice session was frustrating," measured frustration (*M* = 3.0, *SD* = 1.6). Finally, before the posttest, participants provided a judgment of learning ("The instruction I just received prepared me well to answer questions about constraint equations"; *M* = 4.1, *SD* = 1.3) and rated the effectiveness of the feedback they received ("The feedback that I received after each practice problem was helpful"; *M* = 4.6, *SD* = 1.3).

***3.6.3 Performance Measures***

On the pretest and on each posttest question, responses were scored using a preregistered rubric: no credit (0 points), partial credit (1–2 points, or 1–3 on the challenge question), or full credit (3 points, or 4 on the challenge question). Full credit was given for any equation equivalent to the correct answer (see Table 1). Equations that treated the problem as a standard rectangle, ignoring its distinguishing feature (e.g., a missing side, an internal fence, a gate, or a cost structure), earned one point. Equations that reflected the distinguishing feature but contained a single error (e.g., an incorrect total or a missing component) earned two points. On the challenge question, which combined three features, equations that reflected at least two of them earned three points (see the Supplementary Material for the full rubric). Two members of the research team who were blind to condition scored all responses independently (98% agreement), and discrepancies were resolved through discussion.

Pretest scores, expressed as a percentage of the three possible points, were used as a measure of prior knowledge ($M$ = 15%, $SD$ = 30%). Posttest scores, also expressed as percentages of possible points, were analyzed separately for the three practice-aligned questions ($M$ = 46%, $SD$ = 37%), the pretest-aligned question ($M$ = 64%, $SD$ = 46%), and the challenge question that combined multiple learning objectives ($M$ = 41%, $SD$ = 41%).

## 4. Results

We tested our hypotheses with linear regression models, regressing each outcome on three planned orthogonal contrasts for condition. A *Socratic* contrast compared Socratic tutoring (+.667) with explanatory feedback and personalized explanation (−.333 each; correct-answer feedback = 0), weighing the costs and benefits of the Socratic approach (H1a, H1b, and H2). An *Explanation* contrast compared correct-answer feedback (−.75) with the three explanation conditions (+.25 each), testing whether explaining the correct answer helps beyond stating it (H3 and H4). Finally, a *Personalization* contrast, for which we made no directional hypothesis, compared personalized explanation (+.5) with explanatory feedback (−.5; other conditions = 0). Because we were concerned that Socratic tutoring might frustrate participants with little background knowledge, each model also included mean-centered pretest scores and their interactions with the three contrasts. We report effect sizes as standardized mean differences, calculated by dividing each unstandardized contrast estimate by the outcome's residual standard deviation after accounting for the influence of experimental conditions. Full regression output is provided in the Supplementary Material. Data and code to reproduce all analyses are shared at https://github.com/the-oak-lab/scaffolded-error-correction.

### 4.1. Performance Outcomes

**Figure 3**

*Posttest Performance on Practice-Aligned, Pretest-Aligned, and Challenge Questions*

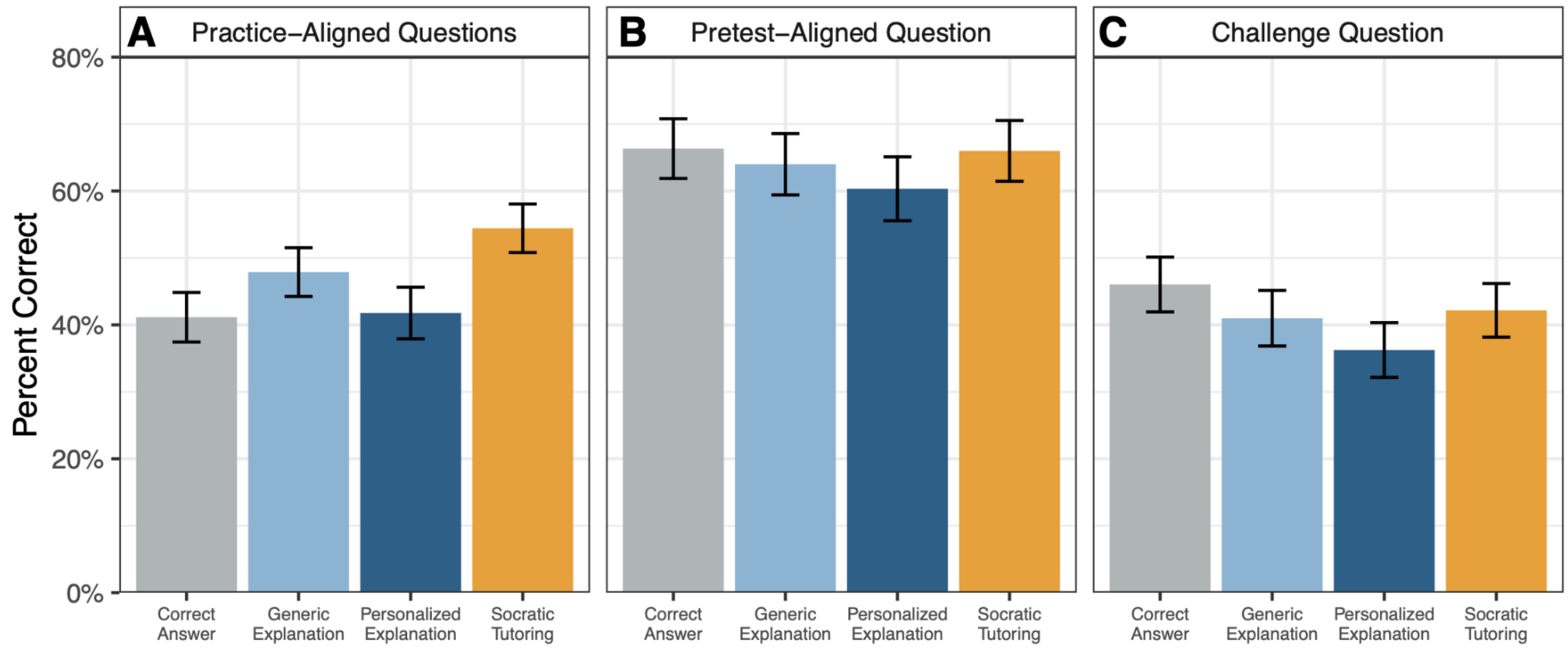

*Note.* Error bars represent +/-1 standard error of the mean.

If the LLM tutor succeeded at diagnosing errors and scaffolding follow-up questions, participants in the Socratic condition should frequently have resolved their errors with its help. They did: the tutor coached Socratic participants to a correct answer after 131 of their 206 errors (64%). Of the 91 Socratic participants who made at least one error, 89% resolved at least one with the tutor's help.

On the posttest, contrary to our preregistered prediction of no benefit (H1a), participants in the Socratic condition scored 9 percentage points (0.24 SD) higher than the average of the explanation-only conditions on the practice-aligned transfer questions, $t(392) = 1.98$, $p = .048$ (Figure 3A).

This advantage did not extend beyond the practiced objectives. Scores on the pretest-aligned question rose substantially, from 15% of possible points on the pretest to 64% on the posttest, even though this objective was not taught during the tutoring session (Figure 3B). This improvement was similar across conditions ($p$s $\geq .603$). Condition effects were also absent on the challenge question, $p$s $\geq .185$ (Figure 3C).

Our hypothesis that explaining the correct answer would help beyond stating it (H3) was not supported. The *Explanation* contrast was not significant on the practice-aligned questions, $t(392) = 1.68$, $p = .094$, the pretest-aligned question, $t(392) = -0.52$, $p = .603$, or the challenge question, $t(392) = -1.33$, $p = .185$. The *Personalization* contrast, which compared personalized explanations with explanatory feedback, was also not significant for any performance outcome, $p$s $\geq .261$.

Pretest scores predicted performance on all three performance outcomes, $p$s $< .001$, but no condition × pretest score interactions emerged, $p$s $\geq .126$.

**4.2 Practice Time**

As predicted (H1b), Socratic tutoring took nearly twice as long as the other conditions: participants in the Socratic condition spent an average of 13.3 minutes on the tutoring session, compared with 6.8 minutes in the correct-answer condition, 6.8 minutes in the generic-explanation condition, and 7.3 minutes in the personalized-explanation condition. The *Socratic* contrast was significant, $t(392) = 10.70$, $p < .001$. The *Explanation* contrast was also significant, $t(392) = 4.22$, $p < .001$, driven primarily by the Socratic condition. Although personalized explanations took about half a minute longer than generic ones, the *Personalization* contrast was not significant, $t(392) = 0.91$, $p = .364$. Pretest scores did not predict practice time, $t(392) = 0.08$, $p = .935$, and no contrast interacted with pretest scores, $p$s $\geq .074$.

## 4.3. Perceived Cost and Frustration

**Figure 4**

*Perceived Cost and Frustration*

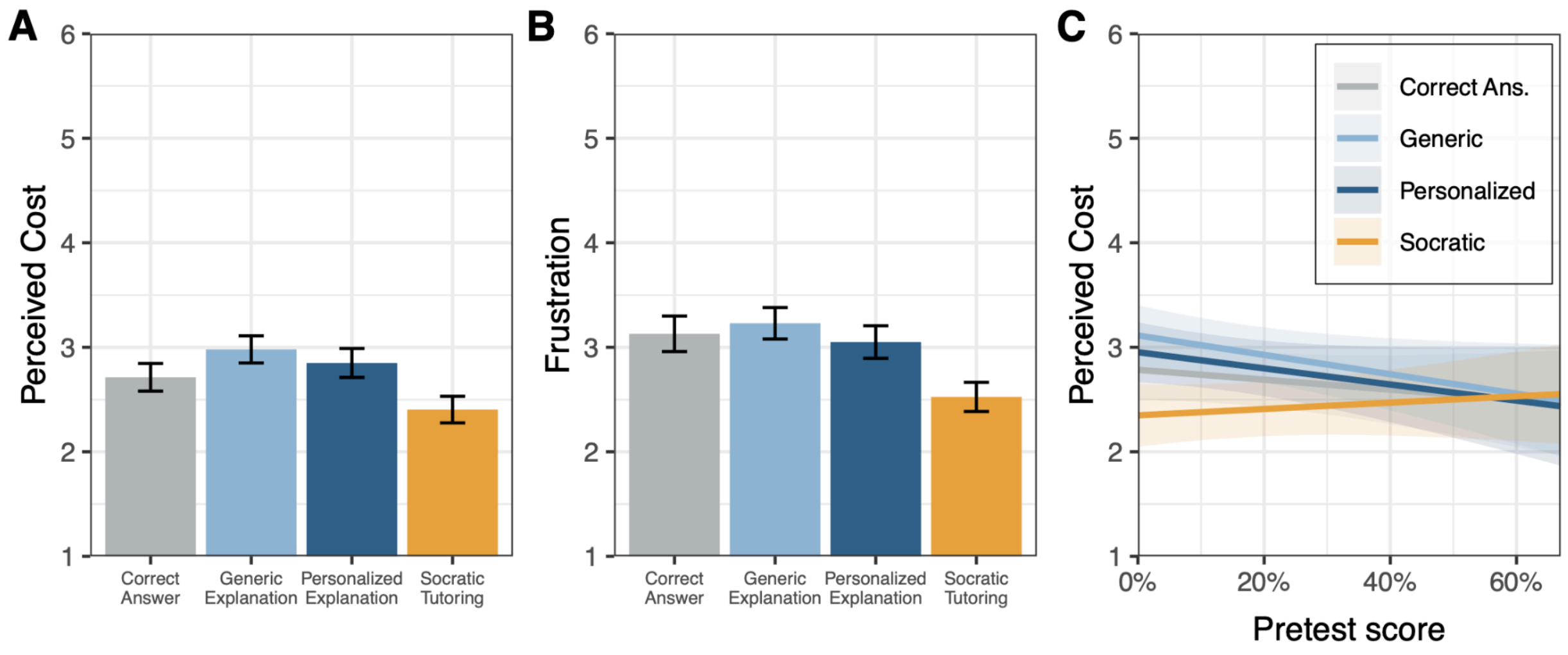

*Note.* Error bars represent +/-1 standard error of the mean.

We next asked whether the extra time Socratic tutoring required made it feel more costly or frustrating. Contrary to our prediction (H2), participants in the Socratic condition reported lower perceived costs than those in the explanation-only conditions (0.38 SD), $t(392) = -3.14$, $p = .002$ (Figure 4A), despite spending nearly twice as long on the practice problems. They also reported less frustration (0.39 SD), $t(392) = -3.17$, $p = .002$ (Figure 4B). Neither the *Explanation* contrast nor the *Personalization* contrast was significant for either outcome ($ps \geq .247$).

The *Socratic* contrast also interacted with pretest scores for perceived costs, $b = 0.35$, $t(392) = 2.18$, $p = .030$, indicating that the cost-related benefit of Socratic tutoring was larger for students with lower pretest scores (Figure 4C). This runs counter to our concern that scaffolded questioning might frustrate students with little background knowledge. A similar but nonsignificant interaction occurred for frustration, $b = 0.34$, $t(392) = 1.84$, $p = .067$.

## 4.4 Interest in the Session and the Topic

**Figure 5**

*Situational Interest in the Learning Session and Interest in Constraint Equations*

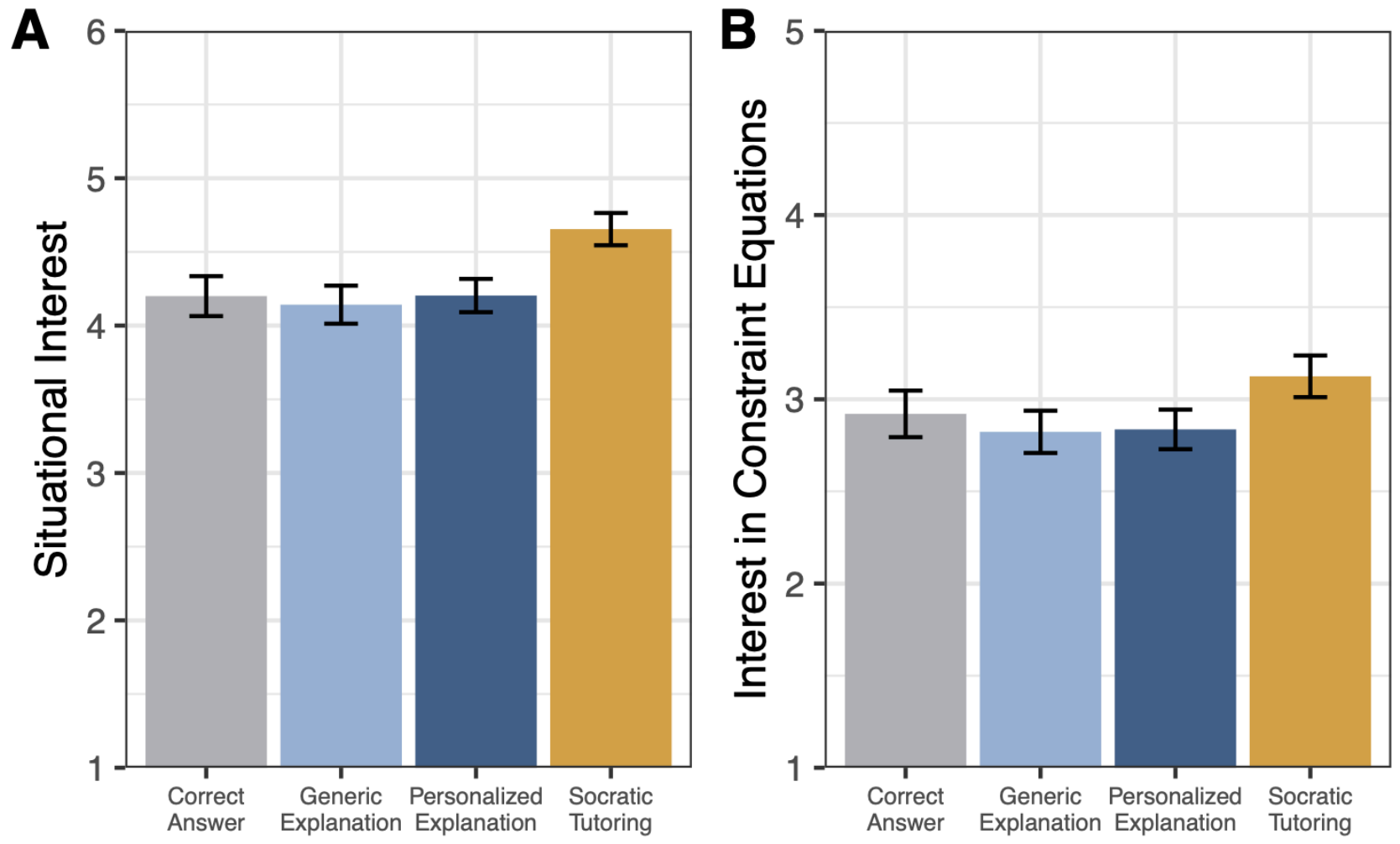

*Note.* Error bars represent +/-1 standard error of the mean.

We next asked whether Socratic tutoring was also more interesting. Indeed, participants in the Socratic condition reported greater situational interest in the session than those in the explanation-only conditions (0.38 SD), $t(392) = 3.08$, $p = .002$ (Figure 5A). Socratic tutoring showed a similar but nonsignificant advantage for interest in constraint equations (0.24 SD; Figure 5B), $t(392) = 1.92$, $p = .055$. Neither the *Explanation* contrast nor the *Personalization* contrast was significant for situational interest ($ps \geq .320$) or for interest in the topic ($ps \geq .856$).

Participants with higher pretest scores reported more situational interest, $b = 0.18$, $t(392) = 2.94$, $p = .003$, and more interest in the topic, $b = 0.15$, $t(392) = 2.59$, $p = .010$. Unlike perceived costs, neither interest outcome showed a condition × pretest interaction, $ps \geq .227$.

**4.5 Judgments of Learning and Perceived Effectiveness**

**Figure 6**

*Judgments of Learning and Perceived Effectiveness of the Feedback*

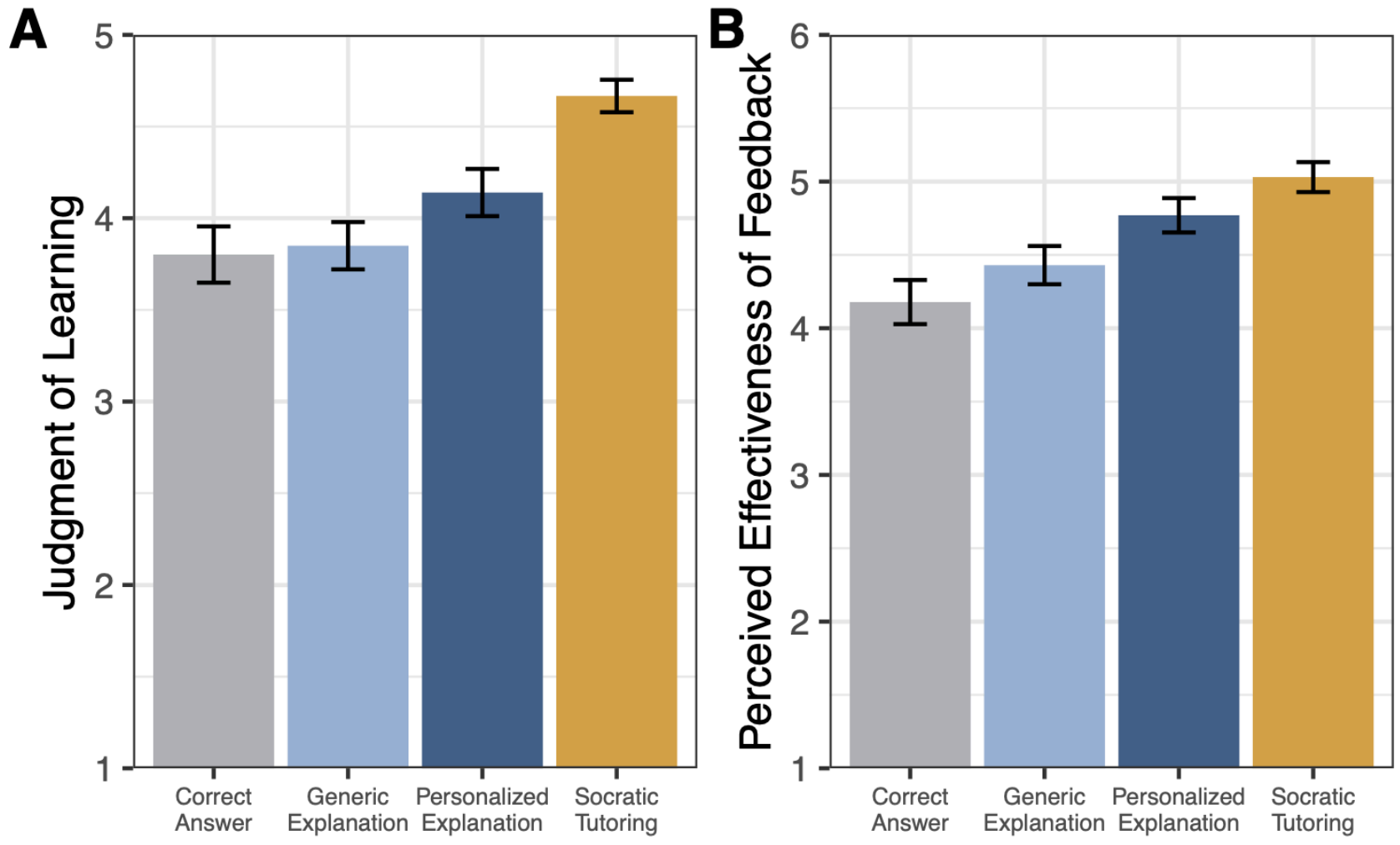

*Note.* Error bars represent +/-1 standard error of the mean.

Finally, we examined how participants judged their own learning and the feedback they received. Participants in the Socratic condition gave higher judgments of learning than those in the explanation-only conditions (0.52 SD), $t(392) = 4.23$, $p < .001$ (Figure 6A), and rated the feedback as more effective (0.34 SD; Figure 6B), $t(392) = 2.73$, $p = .007$.

Consistent with our preregistered prediction (H4), participants in the three explanation conditions perceived the feedback as more effective than participants who received only the correct answer (0.45 SD), $t(392) = 3.91$, $p < .001$. They also gave higher judgments of learning (0.33 SD), $t(392) = 2.89$, $p = .004$. The *Personalization* contrast was not significant for judgments of learning, $t(392) = 1.70$, $p = .091$, or perceived effectiveness, $t(392) = 1.95$, $p = .052$.

Participants with higher pretest scores gave higher judgments of learning, $b = 0.13$, $t(392) = 2.08$, $p = .038$, but pretest scores did not predict perceived effectiveness ($p = .196$). No contrast interacted with pretest scores for either outcome ($ps \geq .234$).

**4.6. Do Resolved Errors Matter?**

Socratic tutoring improved performance on the practice-aligned transfer questions, it increased situational interest and judgments of learning, and it reduced frustration and perceived cost. As we suggested in Section 1.2, these benefits may arise through mastery experiences: when coaching after an error guides a learner to the correct answer, a mistake becomes a success, which should support both learning and motivation. This account predicts that, within the Socratic condition, participants who resolved more of their errors should show better learning and motivation.

To test this prediction, we conducted an exploratory correlational analysis within the Socratic condition ($n$ = 99), testing whether resolved errors were associated with better outcomes than unresolved errors, holding constant the number of first-attempt successes. Figure 7 shows the results of this analysis, comparing the apparent benefits of resolving an error to those of getting a problem correct on the first attempt (in SD units, scaled as in the main analyses).

**Figure 7**

*Associations Between Initial Successes, Resolved Errors, and Outcomes with Socratic Tutoring*

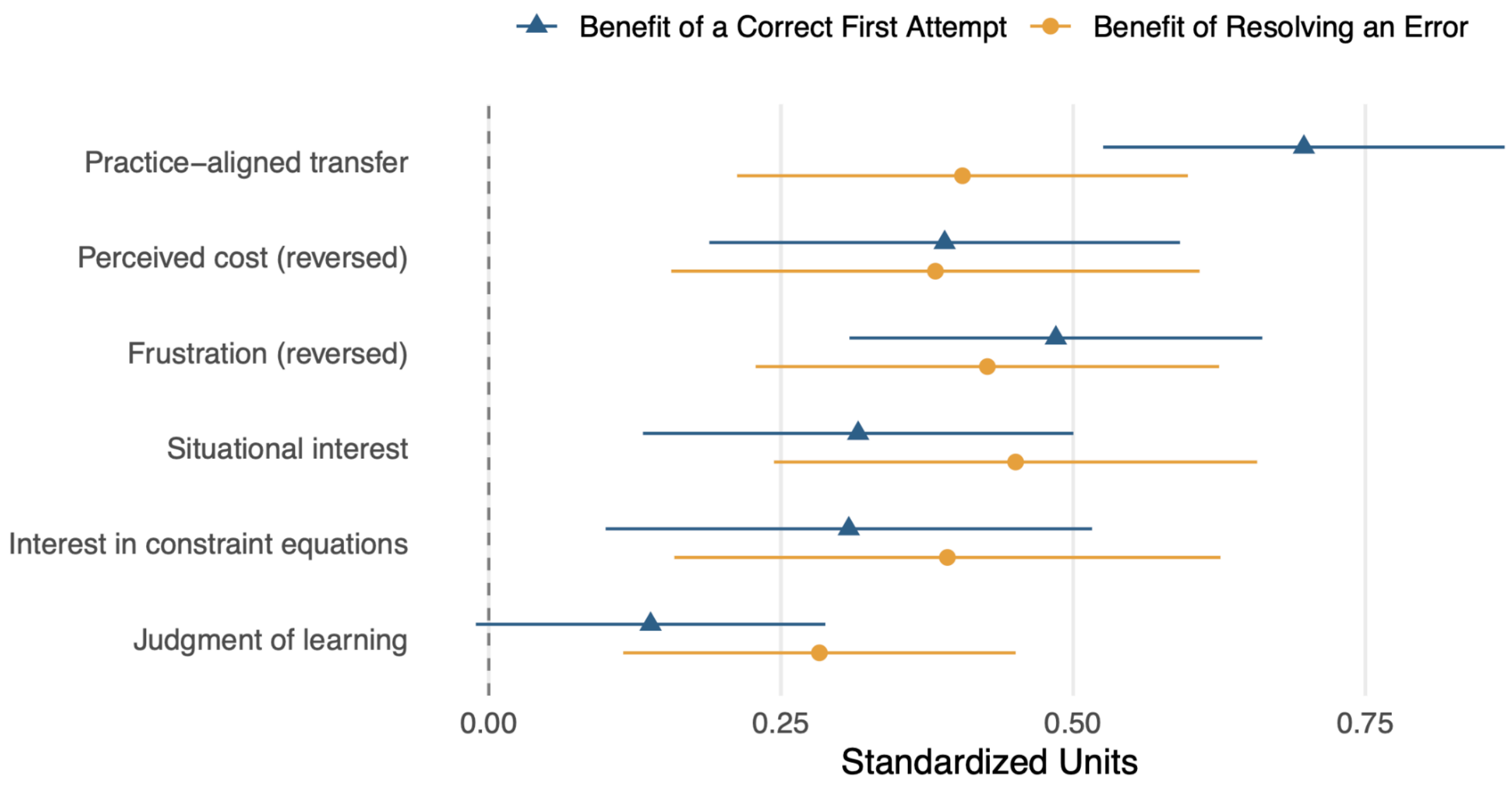

*Note.* Points represent regression coefficients and whiskers represent 95% confidence intervals.

Each resolved error was associated with higher practice-aligned performance, $b = 0.41$, $t(96) = 4.17$, $p < .001$, lower perceived cost, $b = -0.38$, $t(96) = -3.36$, $p = .001$, less frustration, $b = -0.43$, $t(96) = -4.27$, $p < .001$, higher situational interest, $b = 0.45$, $t(96) = 4.33$, $p < .001$, more interest in constraint equations, $b = 0.39$, $t(96) = 3.33$, $p < .001$, and higher judgments of learning, $b = 0.28$, t(96) = 3.35, $p = .001$. In short, to the extent that students were able to resolve their errors, they experienced improved outcomes.

For most outcomes, these effect sizes were similar in magnitude to those of first-attempt successes. When we tested differences between the coefficients, resolved errors had a smaller association with near-transfer performance than first-attempt success did ($b = 0.41$ vs. 0.70), $F(1, 96) = 8.97$, $p = .003$, which likely reflects greater prior knowledge among participants who answered correctly on their own. Descriptively, resolved errors also had a larger association with judgments of learning ($b = 0.28$) than first-attempt successes did ($b = 0.14$), although this difference was not statistically significant, $F(1, 96) = 2.89$, $p = .092$. For every other outcome, the two associations did not reliably differ, $p$s $\geq .201$.

## 5. Discussion

LLMs offer the promise of personalized, adaptive instruction at scale, yet early evaluations of generative AI tutors have found modest benefits at best (Bastani et al., 2025; Oreopoulos & Low, 2026). How can LLM tutoring be designed to support the processes that make one-on-one tutoring effective? Drawing on learning and motivation sciences, we tested two approaches. The first focuses on personalized feedback, using LLMs to explain each student's specific error. The second focuses on scaffolded error correction: a Socratic tutor built into the feedback process steps in after each error and guides students toward the correct response. In an online experiment, we compared these approaches with two benchmarks: correct-answer feedback and pre-written explanatory feedback.

Contrary to our predictions, the additional opportunities created by Socratic tutoring were productive. We expected that they would not justify their costs: that Socratic tutoring would produce no benefit for learning (H1a), take much more time (H1b), and increase the perceived cost of learning (H2). Participants in the Socratic condition did take more time, but they used that time well. The tutor guided participants to resolve 64% of their errors, and 89% of participants who made a mistake resolved at least one error with its help. Participants in the Socratic condition then outperformed those in the explanation-only conditions on questions aligned with the practiced skills (although there were no benefits on questions that involved related skills). Motivational benefits were even more striking: despite the extra time, Socratic participants reported more situational interest, lower perceived costs, and less frustration. These results suggest that when an LLM tutor is a proactive part of the feedback process, rather than an optional guide on the side of the page, it can make practice motivating and effective at building aligned skills.

In contrast, the other three conditions of the study produced very similar outcomes. We found no evidence that adding a personalized explanation to the generic one improved learning. This null result may reflect how the personalized condition was designed. Because it added error-specific information on top of a generic explanation rather than replacing it, the added explanation may have been redundant, or it may have increased the amount students had to process, offsetting any benefit of tailoring.

In addition, and contrary to our prediction, explaining the correct answer did not reliably improve learning over simply showing it (H3), even though participants rated explanations as more effective (H4). This result is surprising. Meta-analytic evidence shows that elaborated feedback typically outperforms correct-answer feedback for both recall and transfer (Mertens et al., 2022). One possible explanation is that we restricted our sample to learners with low prior knowledge. Some prior work suggests that although elaborated explanations can help more advanced learners understand underlying principles, novices may not be equipped to parse them (Asher & Carvalho, 2026). Thus, the novices in our study may have gained as much from comparing their answer with the correct one as from an explanation. For novices, the value of tutoring may lie less in how errors are explained and more in how problems are scaffolded.

### 5.1. A Mastery-Experience Framework of Effective Tutoring

Our findings, together with recent work on practice-centered instruction (Asher et al., 2025; Asher & Carvalho, 2026) and a long history of research on the testing effect, mastery learning, and deliberate practice (e.g., Bloom, 1968; Ericsson et al., 1993; Roediger & Karpicke, 2006), suggest that students learn most effectively when they can complete a mastery-experience loop: attempting a task of appropriate difficulty, making errors that reveal gaps in their understanding, receiving targeted feedback that addresses those gaps, and trying again until they succeed (Figure 8). We propose this is what an effective tutor, AI or human, does. The loop suggests three principles, each supported by existing evidence and each generating predictions about which types of tutor support will be beneficial or harmful.

**Figure 8**

*Conceptual Model of Mastery-Experience Tutoring*

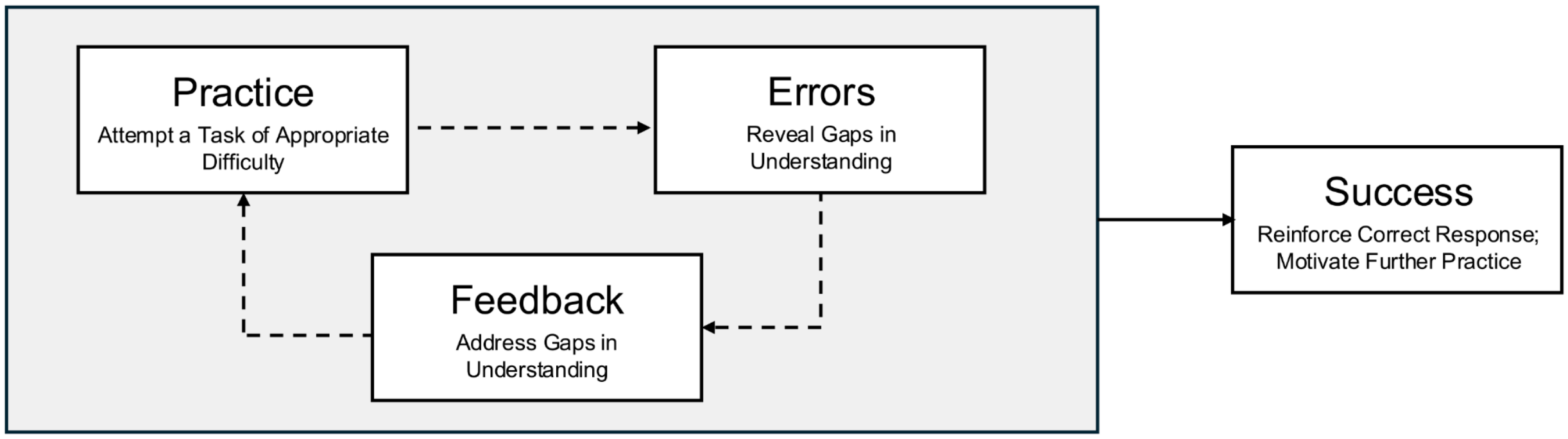

The first principle is that instruction should be centered on practice. In terms of the Knowledge-Learning-Instruction framework (Koedinger et al., 2012), passive instruction (e.g., lecture or reading) consists of an *instructional event* that presents information, whereas feedback is both an *assessment* and an *instructional event*. As a result, passive instruction does little to guide students' attention to relevant information, producing a familiar "in one ear and out the other" feeling (Carvalho et al., 2024; Asher & Carvalho, 2026). With practice and feedback, engagement is built into the task: to attempt a practice problem, a learner must actively reason through the content, in the process reflecting on what they do and do not know. When feedback is delivered, passive review of information should therefore be transformed into an active, goal-directed search for errors and their causes. All conditions in our study were built around upfront practice and feedback, and all produced substantial gains on the pretest-aligned question. In contrast, any tutor that allows a student to seek help before making an attempt should undermine learning. This is what happens when students are given access to off-the-shelf LLMs that reveal answers prematurely (Bastani et al., 2025; Kreijkes et al., 2026; Stadler et al., 2024).

The second principle is that practice should be of appropriate difficulty, so that it produces informative errors without overwhelming working memory. Cognitive load theory holds that students struggle with problems containing many unlearned elements (Kirschner, 2002; Sweller, 2011), and scaffolding, such as breaking a problem into smaller pieces, keeps the load manageable (Van de Pol et al., 2010). The socratic tutor in our study did this, turning problems that a student could not solve into sub-questions that they could. The principle implies that the benefit of scaffolding depends on the match between a problem and learner, and that it should be largest when a problem is too hard to solve unaided but within reach with sub-questions. It predicts little advantage for Socratic tutoring on easy problems, and little benefit for learners who are so far behind that even sub-questions overwhelm them. It also suggests that the value of a skilled tutor may lie in their ability to construct sub-questions for students with sizable gaps in their background knowledge.

The third principle is that students need the chance to try again and succeed. We developed the motivational case for this principle in Section 1.2: unresolved errors can signal inability, whereas overcoming an error should create an efficacy-building mastery experience. Our results are consistent with this principle. The Socratic tutor, the only condition that returned students to each problem after an error and helped them succeed, not only improved performance on the practiced objectives but also increased interest and reduced frustration and perceived cost. Furthermore, within the Socratic condition, errors coached to a correct answer were associated with better learning and motivation than errors left unresolved. This principle and our findings suggest that the next generation of AI tutors should be proactive, stepping in when students struggle and helping them resolve their errors.

### 5.2. Limitations and Next Steps

Several features of our design point to useful next studies. First, our results come from one task and one sample: a single session in which adults recruited online learned to set up constraint equations. It will be important to explicitly test the limits of when Socratic AI tutoring can be helpful by varying the difficulty of the content and learners' background knowledge. We predict that when problems are easy enough for students to solve on their own, non-interactive feedback will suffice and may be more efficient. And when students are so far out of their depth that even sub-questions are overwhelming, we expect that none of the approaches tested here (correct-answer feedback, explanations, or Socratic tutoring) will be productive. Such students may first need extended instruction in prerequisite skills. Studies that test the limits of Socratic tutoring would also benefit from a comparison with skilled human tutoring, which could show how LLMs might improve their questioning to help struggling students succeed.

Second, our conditions compared bundled interventions rather than isolated mechanisms. Socratic tutoring differed from the explanation-only conditions in several ways at once: it asked participants to try again after an error, broke problems into adaptively scaffolded sub-questions, and gave personalized feedback along the way. In addition, it delivered all of this in a brief, conversational, and encouraging tone. Our mastery-experience account attributes the benefits to returning students to practice and scaffolding them toward success, but our design cannot rule out the other features. Future work should vary these components independently to identify which ones drive learning and motivation.

Third, it remains uncertain whether the learning-related benefits of Socratic tutoring are worth its cost in time. Socratic tutoring improved near-transfer learning by .24 SD relative to the explanation-only conditions but took nearly twice as long. Would participants in the other conditions have experienced similar or greater benefits if they spent an equal amount of time completing additional practice problems with feedback? We predict that in a time-equated comparison, scaffolded error correction should remain the better option for students with many gaps in their background knowledge. For these learners, generic explanations could be overwhelming and therefore unproductive, whereas sub-questions and brief explanations should keep cognitive load manageable, promoting learning, success, and motivation.

Finally, our study does not show that an LLM is necessary for scaffolded questions and feedback to deliver these benefits, and LLM tutoring has monetary costs that matter at scale. Our tutor used Gemini 3.5 Flash with default settings, which include a “medium” level of thinking. Each tutor message averaged approximately 800 input tokens and 900 output tokens, at a cost of about one cent (USD), for a total of approximately 12 cents per participant in the Socratic condition. This is small relative to the cost of a human tutor, but it adds up quickly when tutoring is offered to millions of students across many sessions. Two lines of work could reduce these costs. One is to test whether LLMs can tutor effectively without thinking tokens, which would have cut costs by about 80% for this study. The other is to identify which components of tutoring require an LLM at all. Scaffolded sub-questions could be written in advance for common errors, and retry-until-correct practice requires no generative AI. Comparing LLM tutoring with pre-authored versions of the same components would show where LLMs add value. It would also inform hybrid designs, in which pre-written scaffolds handle common errors and LLMs step in only when they are needed.

## 6. Conclusion

LLMs promise personalized, adaptive instruction at scale, yet early AI tutors have delivered modest results. We tested two ways LLMs could support the processes that make tutoring effective: personalized feedback, in which an LLM explained each student’s errors, and scaffolded error correction, in which a Socratic tutor stepped in after each error and guided students toward the correct response. Adding personalized explanations to generic ones did not detectably improve learning, and neither type of explanation reliably outperformed simply showing the correct answer. But when a Socratic LLM tutor was built into the feedback process, guiding students back to the correct answer after each error, it improved learning on the practiced skills, increased interest, and it reduced frustration and perceived cost, even though students spent nearly twice as long practicing. These findings fit a mastery-experience account of effective tutoring. In this account, effective tutoring is proactive and mastery-oriented: the tutor notices when a student is struggling, steps in with guidance, and works with them until the struggle resolves into a mastery experience. AI tutors should be designed with these principles in mind.

**Declaration of generative AI and AI-assisted technologies in the manuscript preparation process.**

During the preparation of this work the authors used Claude Sonnet 5.5 to copy edit the manuscript. After using this tool/service, the authors reviewed and edited the content as needed and take full responsibility for the content of the published article.

**References**

Aleven, V., McLaren, B. M., Sewall, J., Van Velsen, M., Popescu, O., Demi, S., Ringenberg, M., & Koedinger, K. R. (2016). Example-tracing tutors: Intelligent tutor development for non-programmers. *International Journal of Artificial Intelligence in Education*, *26*(1), 224–269. https://doi.org/10.1007/s40593-015-0088-2

Arztmann, M., Domínguez Alfaro, J. L., Hornstra, L., Wong, J., Jeuring, J., & Kester, L. (2025). Game over? Investigating students' working memory, situational interest, and behavioral patterns as predictors of dropout in an educational game. *International Journal of Child-Computer Interaction*, *43*, 100721. https://doi.org/10.1016/j.ijcci.2025.100721

Asher, M. W., & Carvalho, P. F. (2026). Conditions for effective learning without upfront instruction: How practice with feedback supports memory, generalization, motivation, and metacognition. *Educational Psychology Review*, *38*(1), 12. https://doi.org/10.1007/s10648-025-10103-6

Asher, M. W., Gold, G., Chen, E., & Carvalho, P. F. (2026). Chatbots are undermining crowdsourced research in the behavioral sciences: Detecting artificial intelligence–assisted cheating with a keystroke-based Tool. *Advances in Methods and Practices in Psychological Science*, *9*(1), 25152459261424723. https://doi.org/10.1177/25152459261424723

Asher, M. W., & Harackiewicz, J. M. (2024). Using choice and utility value to promote interest: Stimulating situational interest in a lesson and fostering the development of interest in statistics. *Journal of Educational Psychology*, *117*(4). https://doi.org/10.1037/edu0000921

Asher, M. W., Sana, F., Koedinger, K. R., & Carvalho, P. F. (2025). Practice with feedback versus lecture: Consequences for learning, efficiency, and motivation. *Journal of Applied Research in Memory and Cognition*, *14*(3), 355–368. https://doi.org/10.1037/mac0000205

Bandura, A. (1997). *Self-efficacy: The exercise of control.* Freeman.

Bastani, H., Bastani, O., Sungu, A., Ge, H., Kabakcı, Ö., & Mariman, R. (2025). Generative AI without guardrails can harm learning: Evidence from high school mathematics. *Proceedings of the National Academy of Sciences*, *122*(26), e2422633122. https://doi.org/10.1073/pnas.2422633122

Bjork, R. A., Dunlosky, J., & Kornell, N. (2013). Self-regulated learning: Beliefs, techniques, and illusions. *Annual Review of Psychology*, *64*(1), 417–444. https://doi.org/10.1146/annurev-psych-113011-143823

Bloom, B. S. (1968). Learning for mastery. Instruction and curriculum. Regional education laboratory for the Carolinas and Virginia, topical papers and reprints, number 1. *Evaluation Comment*, *1*(2), n2.

Botelho, A. F., Varatharaj, A., Inwegen, E. G. V., & Heffernan, N. T. (2019). Refusing to try: Characterizing early stopout on student assignments. *Proceedings of the 9th International Conference on Learning Analytics & Knowledge*, 391–400. https://doi.org/10.1145/3303772.3303806

Carpenter, S. K., Pan, S. C., & Butler, A. C. (2022). The science of effective learning with spacing and retrieval practice. *Nature Reviews Psychology*, *1*(9), 496–511. https://doi.org/10.1038/s44159-022-00089-1

Carvalho, P. F., Asher, M. W., Sana, F., & Koedinger, K. R. (2024). *Skip the reading assignment: Effective and efficient learning with only practice and feedback*. Open Science Framework. https://doi.org/10.31219/osf.io/9y2tb

Carvalho, P. F., McLaughlin, E. A., & Koedinger, K. R. (2022). Varied practice testing is associated with better learning outcomes in self-regulated online learning. *Journal of Educational Psychology*, *114*(8), 1723–1742. https://doi.org/10.1037/edu0000754

Dermeval, D., Paiva, R., Bittencourt, I. I., Vassileva, J., & Borges, D. (2018). Authoring Tools for Designing Intelligent Tutoring Systems: A Systematic Review of the Literature. *International Journal of Artificial Intelligence in Education*, *28*(3), 336–384. https://doi.org/10.1007/s40593-017-0157-9

Dietrichson, J., Bøg, M., Filges, T., & Klint Jørgensen, A.-M. (2017). Academic interventions for elementary and middle school students with low socioeconomic status: A systematic review and meta-analysis. *Review of Educational Research*, *87*(2), 243–282. https://doi.org/10.3102/0034654316687036

Durik, A. M., Shechter, O. G., Noh, M., Rozek, C. S., & Harackiewicz, J. M. (2015). What if I can't? Success expectancies moderate the effects of utility value information on situational interest and performance. *Motivation and Emotion*, *39*(1), 104–118. https://doi.org/10.1007/s11031-014-9419-0

Eames, T., Brunskill, E., Yamkovenko, B., Weatherholtz, K., & Oreopoulos, P. (2026). Computer-assisted learning in the real world: How Khan Academy influences student math learning. *Proceedings of the National Academy of Sciences*, *123*(1), e2507708123. https://doi.org/10.1073/pnas.2507708123

Ericsson, K. A., Krampe, R. T., & Tesch-Römer, C. (1993). The role of deliberate practice in the acquisition of expert performance. *Psychological Review*, *100*(3), 363–406. https://doi.org/10.1037/0033-295X.100.3.363

Flake, J. K., Barron, K. E., Hulleman, C. S., McCoach, B. D., & Welsh, M. E. (2015). Measuring cost: The forgotten component of expectancy-value theory. *Contemporary Educational Psychology*, *41*, 232–244. https://doi.org/10.1016/j.cedpsych.2015.03.002

Freeman, S., Eddy, S. L., McDonough, M., Smith, M. K., Okoroafor, N., Jordt, H., & Wenderoth, M. P. (2014). Active learning increases student performance in science, engineering, and mathematics. *Proceedings of the National Academy of Sciences*, *111*(23), 8410–8415. https://doi.org/10.1073/pnas.1319030111

Gold, G., Tjaden, J., & Carvalho, P. F. (2026). *Less talk, more code: Practice-based instruction improves programming skill acquisition*. PsyArXiv. https://doi.org/10.31234/osf.io/p428m_v1

Hecht, C. A., Grande, M. R., & Harackiewicz, J. M. (2021). The role of utility value in promoting interest development. *Motivation Science*, *7*(1), 1–20. https://doi.org/10.1037/mot0000182

Johansson, A., Savi, A. O., Van Der Maas, H. L. J., & Hofman, A. D. (2024). *Three strikes and who is out? Individual differences in error-induced quitting*. PsyArXiv. https://doi.org/10.31234/osf.io/6tduv

Kirschner, P. A. (2002). Cognitive load theory: Implications of cognitive load theory on the design of learning. *Learning and Instruction*, *12*(1), 1–10. https://doi.org/10.1016/S0959-4752(01)00014-7

Kirschner, P. A., Sweller, J., & Clark, R. E. (2006). Why minimal guidance during instruction does not work: An analysis of the failure of constructivist, discovery, problem-based, experiential, and inquiry-based teaching. *Educational Psychologist*, *41*(2), 75–86. https://doi.org/10.1207/s15326985ep4102_1

Koedinger, K. R., Corbett, A. T., & Perfetti, C. (2012). The knowledge‐learning‐instruction framework: Bridging the science‐practice chasm to enhance robust student learning. *Cognitive Science*, *36*(5), 757–798. https://doi.org/10.1111/j.1551-6709.2012.01245.x

Koedinger, K. R., Kim, J., Jia, J. Z., McLaughlin, E. A., & Bier, N. L. (2015). Learning is not a spectator sport: Doing is better than watching for learning from a MOOC. *Proceedings of the Second (2015) ACM Conference on Learning @ Scale*, 111–120. https://doi.org/10.1145/2724660.2724681

Kreijkes, P., Kewenig, V., Kuvalja, M., Lee, M., Hofman, J. M., Vitello, S., Sellen, A., Rintel, S., Goldstein, D. G., Rothschild, D., Tankelevitch, L., & Oates, T. (2026). Effects of LLM use and note-taking on reading comprehension and memory: A randomised experiment in secondary schools. *Computers & Education*, *243*, 105514. https://doi.org/10.1016/j.compedu.2025.105514

Liebenow, L. W., Schmidt, F. T. C., Meyer, J., & Fleckenstein, J. (2025). Self-assessment accuracy in the age of artificial Intelligence: Differential effects of LLM-generated feedback. *Computers & Education*, *237*, 105385. https://doi.org/10.1016/j.compedu.2025.105385

Linnenbrink-Garcia, L., Durik, A. M., Conley, A. M., Barron, K. E., Tauer, J. M., Karabenick, S. A., & Harackiewicz, J. M. (2010). Measuring situational interest in academic domains.

*Educational and Psychological Measurement*, *70*(4), 647–671. https://doi.org/10.1177/0013164409355699

Macnamara, B. N., Hambrick, D. Z., & Oswald, F. L. (2014). Deliberate practice and performance in music, games, sports, education, and professions: A meta-analysis. *Psychological Science*, *25*(8), 1608–1618. https://doi.org/10.1177/0956797614535810

Meyer, J., Jansen, T., Schiller, R., Liebenow, L. W., Steinbach, M., Horbach, A., & Fleckenstein, J. (2024). Using LLMs to bring evidence-based feedback into the classroom: AI-generated feedback increases secondary students' text revision, motivation, and positive emotions. *Computers and Education: Artificial Intelligence*, *6*, 100199. https://doi.org/10.1016/j.caeai.2023.100199

Nickow, A., Oreopoulos, P., & Quan, V. (2024). The promise of tutoring for PreK–12 learning: A systematic review and meta-analysis of the experimental evidence. *American Educational Research Journal*, *61*(1), 74–107. https://doi.org/10.3102/00028312231208687

Oreopoulos, P., & Low, N. (2026). *One click away: AI tutoring with Khanmigo in a two-year school experiment* (No. W35620; p. w35620). National Bureau of Economic Research. https://doi.org/10.3386/w35620

Pan, S. C., & Carpenter, S. K. (2023). Prequestioning and pretesting effects: A review of empirical research, theoretical perspectives, and implications for educational practice. *Educational Psychology Review*, *35*(4), 97. https://doi.org/10.1007/s10648-023-09814-5

Roediger, H. L., & Karpicke, J. D. (2006). Test-enhanced learning: Taking memory tests improves long-term retention. *Psychological Science*, *17*(3), 249–255. https://doi.org/10.1111/j.1467-9280.2006.01693.x

Ryan, R. M., & Deci, E. L. (2000). Self-determination theory and the facilitation of intrinsic motivation, social development, and well-being. *American Psychologist*, *55*(1), 68–78. https://doi.org/10.1037/0003-066X.55.1.68

Stadler, M., Bannert, M., & Sailer, M. (2024). Cognitive ease at a cost: LLMs reduce mental effort but compromise depth in student scientific inquiry. *Computers in Human Behavior*, *160*, 108386. https://doi.org/10.1016/j.chb.2024.108386

Sweller, J. (2004). Instructional design consequences of an analogy between evolution by natural selection and human cognitive architecture. *Instructional Science*, *32*(1/2), 9–31. https://doi.org/10.1023/B:TRUC.0000021808.72598.4d

Sweller, J. (2011). Cognitive load theory. In *Psychology of Learning and Motivation* (Vol. 55, pp. 37–76). Elsevier. https://doi.org/10.1016/B978-0-12-387691-1.00002-8

Van de Pol, J., Volman, M., & Beishuizen, J. (2010). Scaffolding in teacher–student interaction: A decade of research. *Educational Psychology Review*, *22*(3), 271–296. https://doi.org/10.1007/s10648-010-9127-6

Van der Kleij, F. M., Feskens, R. C. W., & Eggen, T. J. H. M. (2015). Effects of feedback in a computer-based learning environment on students' learning outcomes: A meta-analysis. *Review of Educational Research*, *85*(4), 475–511. https://doi.org/10.3102/0034654314564881

VanLehn, K., Graesser, A. C., Jackson, G. T., Jordan, P., Olney, A., & Rosé, C. P. (2007). When are tutorial dialogues more effective than reading? *Cognitive Science*, *31*(1), 3–62. https://doi.org/10.1080/03640210709336984

Wigfield, A., & Eccles, J. S. (2000). Expectancy-value theory of achievement motivation. *Contemporary Educational Psychology*, *25*, 68–81. https://doi.org/10.1006/ceps.1999.1015

# Supplementary Material for
Turning Errors into Successes: Designing LLM Tutoring Around Scaffolded Error Correction

Michael W. Asher, Meng Cao, and Paulo F. Carvalho

**Corresponding author:** Michael W. Asher
**Email:** masher@andrew.cmu.edu

**Table of Contents**

Additional supporting information can be found at:
https://github.com/the-oak-lab/scaffolded-error-correction

## Correlations and Descriptive Statistics

**Table S1**

*Zero-Order Correlations and Descriptive Statistics*

| Measure | 1 | 2 | 3 | 4 | 5 | 6 | 7 | 8 | 9 | 10 | 11 | 12 |
|---|---|---|---|---|---|---|---|---|---|---|---|---|
| 1. Pretest Score | | | | | | | | | | | | |
| 2. N Practice Correct | .29 | | | | | | | | | | | |
| 3. Practice-Aligned Perf. | .21 | .58 | | | | | | | | | | |
| 4. Pretest-Aligned Perf. | .21 | .49 | .62 | | | | | | | | | |
| 5. Challenge Perf. | .21 | .44 | .52 | .33 | | | | | | | | |
| 6. SI in Session | .15 | .30 | .24 | .20 | .11 | | | | | | | |
| 7. Interest in Topic | .13 | .23 | .14 | .16 | .07 | .80 | | | | | | |
| 8. Practice Time | .03 | -.13 | -.13 | -.12 | -.17 | .23 | .18 | | | | | |
| 9. JOL | .11 | .34 | .31 | .24 | .17 | .66 | .56 | .20 | | | | |
| 10. Effectiveness | .07 | .25 | .26 | .25 | .12 | .62 | .48 | .20 | .74 | | | |
| 11. Perceived Cost | -.10 | -.32 | -.33 | -.27 | -.20 | -.60 | -.47 | .00 | -.50 | -.44 | | |
| 12. Frustration | -.15 | -.45 | -.35 | -.32 | -.21 | -.66 | -.54 | .00 | -.60 | -.49 | .80 | |
| *M* | 15% | 90% | 46% | 64% | 41% | 4.3 | 2.9 | 8.5 | 4.1 | 4.6 | 2.7 | 3.0 |
| *SD* | 30% | 99% | 37% | 46% | 41% | 1.2 | 1.2 | 5.4 | 1.3 | 1.3 | 1.3 | 1.6 |

*Note.* N Practice Correct = Number of Practice Problems Correct on the First Attempt. Perf. = Performance. SI = Situational Interest. JOL = Judgment of Learning. Practice Time is reported in minutes.

**Table S2**

*Means and Standard Deviations in the correct-answer feedback (n = 101), generic explanatory feedback (n = 100), personalized explanation (n = 100), and Socratic tutoring (n = 99) conditions.*

| | Correct-Answer | | Generic Explanation | | Personalized Explanation | | Socratic Tutoring | |
|---|---|---|---|---|---|---|---|---|
| **Variable** | **M** | **SD** | **M** | **SD** | **M** | **SD** | **M** | **SD** |
| 1. Pretest Score | 0.15 | 0.31 | 0.14 | 0.29 | 0.13 | 0.28 | 0.18 | 0.32 |
| 2. N Practice Correct | 0.81 | 0.98 | 0.93 | 1.03 | 0.96 | 0.97 | 0.92 | 0.99 |
| 3. N Practice Fixed | 0.00 | 0.00 | 0.00 | 0.00 | 0.00 | 0.00 | 1.32 | 0.88 |
| 4. Practice-Aligned Perf. | 0.41 | 0.37 | 0.48 | 0.36 | 0.42 | 0.39 | 0.54 | 0.36 |
| 5. Pretest-Aligned Perf. | 0.66 | 0.45 | 0.64 | 0.46 | 0.60 | 0.48 | 0.66 | 0.45 |
| 6. Challenge Perf. | 0.46 | 0.41 | 0.41 | 0.42 | 0.36 | 0.41 | 0.42 | 0.40 |
| 7. SI in Session | 4.20 | 1.36 | 4.14 | 1.29 | 4.20 | 1.13 | 4.65 | 1.09 |
| 8. Interest in Topic | 2.92 | 1.27 | 2.82 | 1.15 | 2.84 | 1.07 | 3.12 | 1.13 |
| 9. Practice Time | 6.83 | 3.16 | 6.77 | 3.81 | 7.32 | 3.96 | 13.26 | 6.97 |
| 10. JOL | 3.80 | 1.54 | 3.85 | 1.29 | 4.14 | 1.29 | 4.67 | 0.88 |
| 11. Effectiveness | 4.18 | 1.51 | 4.43 | 1.30 | 4.77 | 1.17 | 5.03 | 1.01 |
| 12. Perceived Cost | 2.71 | 1.33 | 2.98 | 1.30 | 2.85 | 1.38 | 2.40 | 1.27 |
| 13. Frustration | 3.13 | 1.71 | 3.23 | 1.50 | 3.05 | 1.56 | 2.53 | 1.39 |

## Regression Output from Preregistered Models

**Table S3**

*Regression Output from Preregistered Models*

| | Practice-Aligned Questions | | | | | Pretest-Aligned Question | | | | |
|---|---|---|---|---|---|---|---|---|---|---|
| **Term** | **b** | **smd** | **se** | **t** | **p** | **b** | **smd** | **se** | **t** | **p** |
| Intercept | 0.46 | | 0.02 | 25.52 | .000 | 0.64 | | 0.02 | 28.51 | .000 |
| Explanation | 0.07 | 0.19 | 0.04 | 1.68 | .094 | -0.03 | -0.06 | 0.05 | -0.52 | .603 |
| Socratic | 0.09 | 0.24 | 0.05 | 1.98 | .048 | 0.03 | 0.06 | 0.06 | 0.47 | .636 |
| Personalization | -0.06 | -0.16 | 0.05 | -1.13 | .261 | -0.03 | -0.07 | 0.06 | -0.50 | .620 |
| Pretest Score (Std.) | 0.08 | | 0.02 | 4.23 | .000 | 0.10 | | 0.02 | 4.35 | .000 |
| Explanation x Pretest | -0.04 | | 0.04 | -0.94 | .348 | 0.02 | | 0.05 | 0.47 | .639 |
| Socratic x Pretest | -0.05 | | 0.04 | -1.21 | .228 | -0.07 | | 0.05 | -1.24 | .217 |
| Personalization x Pretest | 0.01 | | 0.05 | 0.13 | .895 | 0.02 | | 0.07 | 0.23 | .817 |
| | Challenge Question | | | | | Practice Time (Minutes) | | | | |
| **Term** | **b** | **smd** | **se** | **t** | **p** | **b** | **smd** | **se** | **t** | **p** |
| Intercept | 0.42 | | 0.02 | 20.71 | .000 | 8.55 | | 0.24 | 36.30 | .000 |
| Explanation | -0.06 | -0.15 | 0.05 | -1.33 | .185 | 2.29 | 0.49 | 0.54 | 4.22 | .000 |
| Socratic | 0.03 | 0.07 | 0.05 | 0.56 | .578 | 6.21 | 1.32 | 0.58 | 10.70 | .000 |
| Personalization | -0.04 | -0.11 | 0.06 | -0.76 | .446 | 0.61 | 0.13 | 0.67 | 0.91 | .364 |
| Pretest Score (Std.) | 0.09 | | 0.02 | 4.24 | .000 | 0.02 | | 0.24 | 0.08 | .935 |
| Explanation x Pretest | -0.07 | | 0.05 | -1.53 | .126 | -0.04 | | 0.53 | -0.07 | .944 |
| Socratic x Pretest | -0.04 | | 0.05 | -0.89 | .372 | 0.05 | | 0.57 | 0.09 | .932 |
| Personalization x Pretest | 0.03 | | 0.06 | 0.55 | .581 | 1.27 | | 0.71 | 1.79 | .074 |
| | Perceived Cost | | | | | Frustration | | | | |
| **Term** | **b** | **smd** | **se** | **t** | **p** | **b** | **smd** | **se** | **t** | **p** |
| Intercept | 2.73 | | 0.07 | 41.44 | .000 | 2.98 | | 0.08 | 38.86 | .000 |
| Explanation | 0.02 | 0.02 | 0.15 | 0.14 | .888 | -0.20 | -0.13 | 0.18 | -1.16 | .248 |
| Socratic | -0.51 | -0.38 | 0.16 | -3.14 | .002 | -0.60 | -0.39 | 0.19 | -3.17 | .002 |
| Personalization | -0.14 | -0.10 | 0.19 | -0.73 | .465 | -0.19 | -0.13 | 0.22 | -0.89 | .374 |
| Pretest Score (Std.) | -0.14 | | 0.07 | -2.11 | .035 | -0.24 | | 0.08 | -3.10 | .002 |
| Explanation x Pretest | 0.00 | | 0.15 | 0.02 | .983 | 0.04 | | 0.17 | 0.22 | .825 |
| Socratic x Pretest | 0.35 | | 0.16 | 2.18 | .030 | 0.34 | | 0.19 | 1.84 | .067 |
| Personalization x Pretest | 0.05 | | 0.20 | 0.24 | .814 | -0.03 | | 0.23 | -0.13 | .898 |
| | Situational Interest in the Session | | | | | Interest in Constraint Equations | | | | |
| **Term** | **b** | **smd** | **se** | **t** | **p** | **b** | **smd** | **se** | **t** | **p** |
| Intercept | 4.31 | | 0.06 | 70.74 | .000 | 2.93 | | 0.06 | 50.85 | .000 |
| Explanation | 0.14 | 0.11 | 0.14 | 1.00 | .320 | 0.01 | 0.01 | 0.13 | 0.09 | .928 |
| Socratic | 0.46 | 0.38 | 0.15 | 3.08 | .002 | 0.27 | 0.24 | 0.14 | 1.92 | .055 |
| Personalization | 0.08 | 0.06 | 0.17 | 0.43 | .664 | 0.03 | 0.03 | 0.16 | 0.18 | .856 |

| Term | b | smd | se | t | p | b | smd | se | t | p |
|---|---|---|---|---|---|---|---|---|---|---|
| Pretest Score (Std.) | 0.18 | | 0.06 | 2.94 | .003 | 0.15 | | 0.06 | 2.59 | .010 |
| Explanation x Pretest | 0.05 | | 0.14 | 0.35 | .730 | 0.13 | | 0.13 | 0.98 | .329 |
| Socratic x Pretest | -0.18 | | 0.15 | -1.21 | .227 | -0.12 | | 0.14 | -0.87 | .388 |
| Personalization x Pretest | 0.09 | | 0.18 | 0.50 | .617 | 0.19 | | 0.17 | 1.09 | .278 |
| | Judgment of Learning | | | | | Perceived Effectiveness of Feedback | | | | |
| **Term** | **b** | **smd** | **se** | **t** | **p** | **b** | **smd** | **se** | **t** | **p** |
| Intercept | 4.12 | | 0.06 | 64.78 | .000 | 4.61 | | 0.06 | 72.61 | .000 |
| Explanation | 0.42 | 0.33 | 0.15 | 2.90 | .004 | 0.57 | 0.45 | 0.15 | 3.91 | .000 |
| Socratic | 0.66 | 0.52 | 0.16 | 4.23 | .000 | 0.43 | 0.34 | 0.16 | 2.73 | .007 |
| Personalization | 0.31 | 0.24 | 0.18 | 1.70 | .091 | 0.35 | 0.28 | 0.18 | 1.95 | .052 |
| Pretest Score (Std.) | 0.13 | | 0.06 | 2.08 | .038 | 0.08 | | 0.06 | 1.30 | .196 |
| Explanation x Pretest | -0.09 | | 0.14 | -0.60 | .552 | -0.09 | | 0.14 | -0.65 | .514 |
| Socratic x Pretest | -0.18 | | 0.15 | -1.19 | .234 | -0.13 | | 0.15 | -0.84 | .402 |
| Personalization x Pretest | 0.19 | | 0.19 | 1.02 | .309 | 0.15 | | 0.19 | 0.76 | .445 |

*Note.* smd = standardized mean difference. These standardized mean differences were calculated by dividing the raw regression coefficients by each outcome's residual standard deviation after accounting for the effects of experimental manipulations.

## Regression Output From Socratic Condition Analysis

**Table S4**

*Regression Output from Socratic Condition Analysis*

| | Practice-Aligned Questions | | | | Judgment of Learning | | | |
|---|---|---|---|---|---|---|---|---|
| **Term** | **b** | **se** | **t** | **p** | **b** | **se** | **t** | **p** |
| Intercept | -0.33 | 0.15 | -2.18 | .032 | 0.06 | 0.13 | 0.45 | .654 |
| N Questions Fixed | 0.41 | 0.10 | 4.17 | .000 | 0.28 | 0.09 | 3.35 | .001 |
| N Correct First Attempts (Centered) | 0.70 | 0.09 | 8.06 | .000 | 0.14 | 0.08 | 1.84 | .069 |
| | Situational Interest in Session | | | | Interest in Constraint Equations | | | |
| **Term** | **b** | **se** | **t** | **p** | **b** | **se** | **t** | **p** |
| Intercept | -0.31 | 0.16 | -1.93 | .056 | -0.35 | 0.18 | -1.94 | .056 |
| N Questions Fixed | 0.45 | 0.10 | 4.33 | .000 | 0.39 | 0.12 | 3.33 | .001 |
| N Correct First Attempts (Centered) | 0.32 | 0.09 | 3.41 | .001 | 0.31 | 0.11 | 2.94 | .004 |
| | Perceived Cost | | | | Frustration | | | |
| **Term** | **b** | **se** | **t** | **p** | **b** | **se** | **t** | **p** |
| Intercept | 0.26 | 0.18 | 1.47 | .144 | 0.27 | 0.15 | 1.78 | .079 |
| N Questions Fixed | -0.38 | 0.11 | -3.36 | .001 | -0.43 | 0.10 | -4.27 | .000 |
| N Correct First Attempts (Centered) | -0.39 | 0.10 | -3.85 | .000 | -0.49 | 0.09 | -5.45 | .000 |

*Note.* Each outcome was standardized using its residual standard deviation after accounting for the effects of the experimental manipulations.

## Description of Additional Measures for Exploratory Analysis

The following exploratory measures collected but were not part of the preregistered tests. They are included in the shared data.

**Baseline measures.** Baseline mathematics confidence was assessed with three items on a five-point scale (e.g., “How good are you at mathematics?”; $\alpha$ = .94, *M* = 2.4, *SD* = 1.0). Baseline interest in mathematics was assessed with three items, also on a five-point scale (e.g., “How interesting do you find mathematics?”; $\alpha$ = .95, *M* = 2.8, *SD* = 1.1). Participants also predicted how well they would do on the posttest if they took it now, before any instruction (a whole number from 0 to 100; *M* = 40.3, *SD* = 29.4), and rated their confidence in that estimate on a single five-point item (*M* = 3.0, *SD* = 1.1).

**Outcome Measures.** Two six-point items assessed distraction during the session (“My mind wandered as I learned about constraint equations today” and “I got distracted as I learned about constraint equations today”; $\alpha$ = .88, *M* = 2,1, *SD* = 1.1). Before the posttest, participants predicted their score again on a 0-100 scale (*M* = 47.2, *SD* = 31.3) and provided another rating of their confidence in that estimate with a single five-point item (*M* = 3.2, *SD* = 1.1).

## LLM Tutoring Prompts

### Socratic Tutoring Prompts

Each call to the large language model (LLM; Gemini 3.5 Flash) sent a single text string made of three parts: (1) one of three instruction blocks (below); (2) the conversation so far; and (3) a closing line. Parts 2 and 3 had the same format in every call, although the conversation in Part 2 grew with each exchange. Part 1 was chosen automatically by custom JavaScript running in Qualtrics. It depended on whether the participant’s most recent equation was correct (as scored by the JavaScript, not the LLM) and, if it was incorrect, on how many tutor replies had already been given. The participant’s submitted equation received the first tutor reply. Once six replies had been given without a correct answer, the instructions told the tutor to state the correct equation and explain it (replies 7 and 8; the chat ended after reply 8).

#### *Part 1A. Student’s answer incorrect, tutor replies 1–6*

You are a Socratic tutor helping a student with a math problem. The student was asked:

"{PROBLEM TEXT}”

The student's answer is INCORRECT or INCOMPLETE. The correct answer is: [CORRECT ANSWER].

Your role is to use Socratic questioning to guide them to the correct answer through a series of scaffolded questions. Focus on the underlying reasoning: {CORRECT EXPLANATION}

Be encouraging, patient, and focus on asking guiding questions rather than giving direct answers. Keep responses concise (2-4 sentences). Use clear, simple language. Do not give the answer directly yet - continue with Socratic questions.

A labeled diagram of the scenario is available. If, based on the student's responses so far, you judge they are specifically struggling to visualize the geometry (not just the algebra) and a diagram would genuinely help right now, include the exact text [SHOW_IMAGE] by itself anywhere in your response -- it will be shown to the student and removed from your message before they see it. Use this sparingly and only when it would truly help: showing it too early undercuts the point of working through the questions above, so prefer guiding questions first and reach for it only if the student seems stuck on the shape itself.

***Part 1B. Student's answer incorrect, tutor replies 7–8***

You are a Socratic tutor helping a student with a math problem. The student was asked:

"{PROBLEM TEXT}"

The student has been working on this for {INTERACTION COUNT} interactions and is still struggling. It's time to be more direct and helpful.

Your role now is to:
1. Acknowledge their effort and reassure them this is challenging
2. PROVIDE THE CORRECT ANSWER: {CORRECT ANSWER}
3. Explain step-by-step WHY this is the answer: {CORRECT EXPLANATION}
4. Ask if they understand and if they have any questions

Be warm, encouraging, and clear. Keep it concise (4-6 sentences).

***Part 1C. Student's answer correct***

You are a Socratic tutor helping a student with a math problem. The student was asked:

"{PROBLEM TEXT}"

The student's answer is CORRECT! They have successfully identified a valid equation: {CORRECT ANSWER DISPLAY}.

Your role is to:
1. Congratulate them enthusiastically and confirm their answer is correct
2. Briefly explain WHY their answer is correct: {CORRECT EXPLANATION}

Keep your response warm, encouraging, and concise (2-3 sentences). End there -- do not ask a follow-up question, invite further discussion, or suggest extending the problem (e.g. into optimization or other extensions). The interface already gives the student a clear way to continue chatting if they want to; your response should read as complete on its own.

***Part 2: Conversation so far***

Conversation so far:

Student: Here is my answer to the problem: {STUDENT'S ANSWER}
Tutor: {TUTOR REPLY 1}
Student: {STUDENT MESSAGE 1}
Tutor: {TUTOR REPLY 2}
...

***Part 3: Closing line***

Respond as the Tutor to the Student's most recent message above.

## Posttest Questions

### Test Question 1: Brick Wall Garden Problem

A homeowner has 240 feet of fencing to create a rectangular garden next to a long brick wall. The side of the garden along the wall does not need fencing.

*Write an equation using x and y for the dimensions of the garden, given the fencing constraint.*

### Test Question 2: Community Garden Problem

A community garden is using 1,800 feet of fencing to enclose a rectangular area. Two internal fences, one running lengthwise and one running widthwise, divide the garden into 4 equal rectangular plots.

*Write an equation using x and y for the dimensions of the entire garden, given the fencing constraint.*

### Test Question 3: Wildlife Enclosure Problem

A wildlife center is building a rectangular outdoor enclosure. One pair of opposite sides will use reinforced fencing costing $8 per foot. Of the other two sides, one will use standard fencing costing $4 per foot, and the other will use secure fencing costing $6 per foot. The total fencing budget is $900.

*Write an equation using x and y for the dimensions of the enclosure, given the budget constraint.*

### Test Question 4: Farm Gates Problem

A farm is fencing a rectangular enclosure using 500 feet of fencing. One side of length x has a 12-foot gate for tractors, and one side of length y has an 8-foot gate for workers.

*Write an equation using x and y for the dimensions of the enclosure, given the fencing constraint.*

### Test Question 5: Greenhouse Corner Problem

A plant nursery is building a rectangular greenhouse space against the corner of a brick building, so two adjacent sides require no fencing. The remaining side of length x uses reinforced decorative steel fencing costing $15 per foot, but includes an 8-foot sliding gate that requires no

fencing. The remaining side of length y uses standard mesh fencing costing $5 per foot. The total fencing budget is $900.

*Write an equation using x and y for the dimensions of the greenhouse, given the budget constraint.*

**Rubric For Scoring Constraint Equation Questions**

**Table S5**

*Rubric for Scoring Constraint Equation Questions*

| **Item** | **Scoring criteria** |
|---|---|
| **Pretest Question: River Fence Problem**<br><br>***Objective: Missing Side***<br><br>A farmer has 200 feet of fencing and wants to build a rectangular enclosure along a straight river. The river side doesn't need fencing.<br><br>Write an equation using x and y for the dimensions of the enclosure, given the fencing constraint. | 3 points: An equation equivalent to $2x + y = 200$ (or $2y + x = 200$)<br><br>2 points: An equation that acknowledges the problem's specific learning objective (sums 3 sides, not 4) but states $2x + y =$ incorrect total.<br><br>1 point: $2x + 2y = 200$: correct rectangle, ignores missing river side. |
| **Posttest Question 1: Brick Wall Garden Problem**<br><br>***Far Transfer Objective (not taught during instruction): Fence with a Missing Side***<br><br>A homeowner has 240 feet of fencing to create a rectangular garden next to a long brick wall. The side of the garden along the wall does not need fencing. | 3 points: An equation that is equivalent to $240 = 2x + y$ (or $240 = 2y + x$).<br><br>2 points: An equation that acknowledges the problem's specific learning objective (sums 3 sides, not 4) but that states $2y + x =$ wrong total (or $2x + y =$ wrong total).<br><br>1 point: $2x + 2y = 240$: correct rectangle, ignores the missing wall |
| Write an equation using x and y for the dimensions of the garden, given the fencing constraint. | |
| Posttest Question 2: Community Garden Problem<br><br>Near Transfer Objective (taught in a simpler form during instruction): Divided Fence<br><br>A community garden is using 1,800 feet of fencing to enclose a rectangular area. Two internal fences, one running lengthwise and one running widthwise, divide the garden into 4 equal rectangular plots.<br><br>Write an equation using x and y for the dimensions of the entire garden, given the fencing constraint. | 3 points: An equation that is equivalent to $3x + 3y = 1800$.<br><br>2 points: An equation that acknowledges the problem's specific learning objective (including an extra fence) but misses the other fence (e.g., $3x + 2y = 1800$). OR that states $3x + 3y =$ incorrect total.<br><br>1 point: $2x + 2y = 1800$: correct perimeter, ignores both internal fences. |
| **Posttest Question 3: Wildlife Enclosure Problem**<br><br>**Near Transfer Objective (taught in a simpler form during instruction): Fence with Material Costs**<br><br>A wildlife center is building a rectangular outdoor enclosure. One pair of opposite sides will use reinforced fencing costing $8 per foot. Of the other two sides, one will use standard fencing costing $4 per foot, and the other will use secure fencing costing $6 per foot. The total fencing budget is $900.<br><br>Write an equation using x and y for the dimensions of the enclosure, given the budget constraint. | 3 points: An equation that is equivalent to $16x + 10y = 900$ (or $16y + 10x = 900$).<br><br>2 points: An equation that acknowledges the problem's specific learning objective (setting up a cost equation) with correct structure but one error e.g., $8x + 10y = 900$ (counting only one reinforced x-side instead of both); OR that states $16x + 10y =$ incorrect total (or $16y + 10x =$ incorrect total).<br><br>1 point: $2x + 2y = 900$: mistakes cost for fence length. |
| Posttest Question 4: Farm Gates Problem<br><br>***Near Transfer Objective (taught in a simpler form during instruction): Fence with a Gap***<br><br>A farm is fencing a rectangular enclosure using 500 feet of fencing. One side of length x has a 12-foot gate for tractors, and one side of length y has an 8-foot gate for workers.<br><br>Write an equation using x and y for the dimensions of the enclosure, given the fencing constraint. | 3 points: An equation that is equivalent to $2x + 2y - 20 = 500$.<br><br>2 points: An equation that acknowledges the problem's specific learning objective (the gap in the fence), but that forgets to subtract one of the two gates (e.g., $2x + 2y - 8 = 500$). OR that states $2x + 2y - 20 =$ incorrect value.<br><br>1 point: $2x + 2y = 500$: correct perimeter, ignores both gates. |
| Posttest Question 5: Greenhouse Corner Problem<br><br>***Far Transfer, Challenge Objective: Includes multiple objectives (Fence with a Gap, Missing Sides, and Costs), including one (Missing Sides) not taught during instruction***<br><br>A plant nursery is building a rectangular greenhouse space against the corner of a brick building, so two adjacent sides require no fencing. The remaining side of length x uses reinforced decorative steel fencing costing $15 per foot, but includes an 8-foot sliding gate that requires no fencing. The remaining side of length y uses standard mesh fencing costing $5 per foot. The total fencing budget is $900.<br><br>Write an equation using x and y for the dimensions of the greenhouse, given the budget constraint. | 4 points: An equation that is equivalent to $15(x - 8) + 5y = 900$.<br><br>3 points: An equation that acknowledges at least 2 of the three learning objectives (e.g., accounting for the corner and costs, but not the gap) but that makes a mistake on the other; OR that states $15(x - 8) + 5y =$ incorrect value.<br><br>2 points: An equation that is set equal to 900 and that acknowledges a gap, missing sides, or costs, but that is otherwise mistaken.<br><br>1 point: $2x + 2y = 900$. A perimeter equation that mistakes cost for fence length, ignores the gap, and ignores missing sides. |